\documentclass[%
 reprint,
 amsmath,amssymb,
 aps,11pt,onecolumn]{revtex4-1}
\usepackage{subfig}
\usepackage{graphicx}
\usepackage{float}
\usepackage{dcolumn}
\pdfoutput=1
\usepackage{setspace}
\usepackage{bm}
\usepackage{sistyle}
\usepackage{braket}
\usepackage{lipsum}
\usepackage{titlesec}
\usepackage{color}
\usepackage[]{subfig}

\begin{document}
\title{ Dependence of the magnetic interactions in MoS$_2$ monolayer on Mn-doping configurations}

\author{ {\small Adlen Smiri}}
\affiliation{{\footnotesize  Facult\'e des Sciences de Bizerte, Laboratoire de Physique des Mat\'eriaux: Structure et Propri\'et\'es,\\
Universit\'e  de Carthage, 7021 Jarzouna, Tunisia}}%
\affiliation{ {\footnotesize 
LPCNO, Universit\'e F\'ed\'erale de Toulouse Midi-Pyr\'en\'ees,\\
INSA-CNRS-UPS, 135 Av. de Rangueil, 31077 Toulouse, France
}}%
\author{{\small Iann C. Gerber}}
\affiliation{ {\footnotesize 
LPCNO, Universit\'e F\'ed\'erale de Toulouse Midi-Pyr\'en\'ees,\\
INSA-CNRS-UPS, 135 Av. de Rangueil, 31077 Toulouse, France
}}%
\author{ \small {Samir Lounis}}
\affiliation{{\footnotesize Peter Gr\"unberg Institut and Institute for
Advanced Simulation, Forschungszentrum J\"ulich and JARA, 52425 J\"ulich, Germany
}}%
\author{{\small Sihem Jaziri}}
\affiliation{{\footnotesize Facult\'e des Sciences de Bizerte, Laboratoire de Physique des Mat\'eriaux: Structure et Propri\'et\'es,\\
Universit\'e  de Carthage, 7021 Jarzouna, Tunisia
}}%
\affiliation{
{\footnotesize Facult\'e des Sciences de Tunis, Laboratoire de Physique de la Mati\'ere Condens\'ee,\\
D\'epartement de Physique, Universit\'e  Tunis el Manar, Campus Universitaire 2092 Tunis, Tunisia
}}%
\begin{abstract}
{ \small
 Understanding the magnetic properties of the various Mn doping configurations that can be encountered in $2H$-MoS$_2$ monolayer could be beneficial for its use in spintronics. Using density functional theory plus Hubbard term (DFT$+$U) approach, we study how a single isolated, double- and triple-substitution configurations of Mn atoms within a MoS$_2$ monolayer could contribute to its total magnetization. We find that the doping-configuration plays a critical role in stabilizing a ferromagnetic state in a Mn-doped MoS$_2$ monolayer. Indeed, the Mn-Mn magnetic interaction is found to be ferromagnetic and strong for Mn in equidistant substitution positions where the separation average range of 6-11 $\angstrom$. The strongest ferromagnetic interaction is found when substitutions are in second NN Mo-sites of the armchair chain. Clustering is energetically favorable but it strongly reduces the ferromagnetic exchange energies. Our results suggest that ordering the Mn dopants on MoS$_2$ monolayer is needed to increase its potential ferromagnetism.
 }
\end{abstract}

\maketitle
\section{Introduction}
On the grounds of their special structural and electronic properties~\cite{ref01,ref02,ref03,ref12,ref13}, the two-dimensional transition metal (M) dichalcogenide (X) semiconductors (2D-TMDs) have shown peculiar optical characteristics~\cite{ref010,ref011,ref012,ref013,ref014} leading to several applications, such as optoelectronics~\cite{ref05,ref07}, including lasers and light-emitting diodes~\cite{ref017,ref018,ref019,ref020}.
In the past few years, doping-induced magnetism in nonmagnetic 2D-TMDs,  such as MoS$_2$~\cite{ref6,ref7,ref8,ref9,ref10,ref14,ref017,ref047,ref060}, WS$_2$~\cite{ref048}, WSe$_2$~\cite{ref048} or SnS$_2$~\cite{ref015} systems has deserved considerable attention. The magnetic properties of doped 2D-TMDs, such as strong ferromagnetism (FM)~\cite{ref6,ref7,ref8,ref9,ref10,ref14,ref15,ref015,ref016,ref060} and large magnetic anisotropy, are sought to be used in the ultimately small magnetic devices. To achieve this purpose, nonmetal (H, B, Cr, etc.)~\cite{ref060} and transition-metal (Mn, Fe, Nb, etc.)~\cite{ref6,ref7,ref8,ref9,ref10,ref14,ref017,ref047,ref060} dopants through various doping strategies, such as substitution at M or X-sites~\cite{ref6,ref7,ref8,ref9,ref10,ref14,ref15,ref015,ref016,ref048} and adsorption \cite{ref047}, have been used to tune magnetism in 2D-TMDs~\cite{ref6,ref7,ref8,ref9,ref10,ref14,ref15,ref015,ref016,ref047,ref048}.\\

\par 
 Among 2D-TMDs, $2H$-MoS$_2$ monolayer (ML) has shown specific electronic transport properties, like considerable electron mobility (up to 1000 cm$^2/Vs$ at low temperature)~\cite{ref049,ref050,ref051,ref054} or low power dissipation~\cite{ref051,ref052,ref053}. These features make this material a promising 2D-TMDs candidate for electronic transport devices, essentially for next-generation transistors~\cite{ref051,ref052,ref053,ref055}. These transport abilities has led to extensive efforts to induce magnetism in MoS$_2$  ML~\cite{ref7,ref8,ref9,ref14,ref017} in order to control the electron spin and thus reach spintronic applications. To this end, one of the effective tools is to substitute some Mo atoms in ML by Manganese ones to produce Mn-doped MoS$_2$, which has attracted wide interest, either theoretically~\cite{ref7,ref8,ref9} or experimentally~\cite{ref14,ref017}. Indeed, using first-principles calculations, it was demonstrated that this substitution is energetically favorable under S-rich regime, which is common in reaction medium for MoS$_2$ nanosheets' synthesis~\cite{ref8}. Moreover, substitutional Mn atoms was found to prefer clustering in MoS$_2$ ML~\cite{ref10}. Experimentally, doped Mn-MoS$_2$ sheets have been successfully synthesized through different methods. In particular, Kehao et \textit{al.}~\cite{ref010}, succeeded to incorporate Mn in MoS$_2$ ML via vapor phase deposition techniques. More recently, a hydrothermal method for Mn-doped MoS$_2$ ML synthesis has been proposed by Jieqiong et \textit{al.}~\cite{ref14}.\\

\par
According to several works~\cite{ref9,ref7,ref10,ref15}, the Mn impurities within MoS$_2$ ML are coupled ferromagnetically. In the earliest study of  Mn-doped MoS$_2$ ML, Ramasubramaniam and Naveh~\cite{ref9} attributed the origin of this FM coupling to the double-exchange magnetic interaction~\cite{ref033}. This type of interaction is due to the presence of delocalized carriers between Mn impurities. However, the double-exchange mechanism was ruled out by Mishra et \textit{al.}~\cite{ref7} based on the fact that there exists an antiferromagnetic (AFM) coupling between Mn atoms and their closest Sulfur atoms. More recently, another origin of magnetic interaction among Mn impurities in MoS$_2$ ML has been proposed and called successive spin polarizations (SSP)~\cite{ref032,ref034}. The SSP magnetic coupling model is based on the spin-polarization induced by impurities in the host material to their nearest environment, i.e the closest atoms mainly~\cite{ref032,ref034,ref035,ref036}. In particular, a specific Mn impurity dictates the spin polarization of its first Next Nearest neighbors (NN), namely Mo and S, which in return dictate the spin polarization of the NN possible dopant~\cite{ref032,ref034}. Unlike the double-exchange mechanism, the SSP FM coupling is based on localized electronic processes that take place between Mn impurities~\cite{ref032}. Therefore, the SSP FM coupling can take place at low magnetic dopant concentrations which can be below the percolation threshold~\cite{ref032,ref034}. Hence, a local enhancement of FM coupling by manipulating Mn-doping configurations may lead to avoid the need of high-doping concentration and still get strong ferromagnetism. Since the doping concentration cannot be easily controlled in experiments~\cite{ref14}, the risk is high to lose the semi-conducting property of MoS$_2$ ML at large dopant concentration~\cite{ref14}. To do this, one must first understand the role of doping configurations in stabilizing the FM state of  MoS$_2$ ML.\\

In both studies of Ramasubramaniam and Naveh~\cite{ref9} and Mishra \textit{et al.}~\cite{ref7}, the FM coupling strength between two Mn impurities was found to decrease with respect to Mn-Mn distance's increasing. Additionally, according to the SSP model, one can expect that the strength of the FM coupling can also depend drastically of the very local configuration between the two Mn atoms~\cite{ref032,ref034,ref035,ref036}. For instance, in the case of \textit{1T}$^{'}-$MoS$_2$ ML doped by substitution with Mn atoms, the strength of their magnetic interaction was found to be highly dependent on their relative positions~\cite{ref056}. Indeed, the FM coupling was found more pronounced when two Mn dopants were separated by 6.38 $\angstrom$ than by 3.81 $\angstrom$~\cite{ref056}. Mind that similar conclusions have been drawn already in the case of FM coupling between Co atoms embedded in a single graphene sheet~\cite{ref057}.\\

In Ref.~\cite{ref9} it was shown that the ferromagnetism in Mn-doped MoS$_2$ ML becomes important when the Mn-doping concentration increases. In particular, in 10-15\% Mn-doping range, the Curie temperature was found to be above room temperature. Motivated by this result, Jieqiong et \textit{al.}~\cite{ref14} succeeded to elaborate a MoS$_2$ ML heavily doped with Mn impurities which gives rise to robust ferromagnetism. However, they also demonstrated that the different resulting doping configurations contribute differently,  and even not, to the overall ML's ferromagnetism~\cite{ref14}. On the one hand, those Mn with NN forming Mn clusters are typically antiferromagnetic and thus do not contribute to the overall magnetization. On the other hand, only those Mn dopants that are at suitable distances can order ferromagnetically~\cite{ref14}. The diversity of magnetic behaviour of the different Mn doping configurations in MoS$_2$ ML results on two different FM phases in this material~\cite{ref14}. Therefore, distinguishing these different contributions is of high interest in order to potentially control magnetism in Mn-doped MoS$_2$ ML.\\

Using Density Functional Theory plus Hubbard term (DFT$+$U), we perform a comprehensive investigation of structural stability and magnetic properties, namely magnetic exchange interaction and magnetic moments, of few near Mn-dopants embedded in MoS$_2$ ML. By placing  Mn atoms in armchair- and/or zigzag-substitution Mo-sites with different Mn-Mn distances, we generate several doping configurations. Our aim is to explore the effect of these doping configurations on the magnetic coupling nature and strength among Mn impurities. To this end, the outline of this paper is as follows: we start by a description of our computational details and methods in section $\textbf{II}$. In section $\textbf{III}$ we present and discuss our results for MoS$_2$ ML with multiple Mn dopings: in $\textbf{III. A}$, we validate our  computational parameters and approachs by comparing our results of the isolated Mn-induced magnetic and electronic properties to that of literature. In $\textbf{III. B}$, we study the structural stability, pairwise exchange interaction of two Mn-dopants placed on armchair or zigzag chains as a function of Mn-Mn separations. In $\textbf{III. C}$, to broaden our understanding of the magnetic exchange interaction behavior versus the doping configurations, we add a third Mn-dopant. Indeed, by manipulating the three dopant positions, we are able to determine the effect of the doping clustering, doping shape (tiangle- or line- like) and equidistant or non-equidistant doping on the magnetic properties. We also discuss the dependence of the magnetic exchange interaction on inter-dopant distances. Our results are compared to previous calculations and to the experiment of Jieqiong et \textit{al.}~\cite{ref14}. Finally, we conclude our results in section  $\textbf{IV}$.
\section{Methods and Computational details}
 Our  work was based on spin-polarized DFT implemented in Vienna \textit{ab initio} simulation package (VASP)~\cite{ref1,ref2}. The exchange-correlation interaction was described using the  Perdew-Burke-Ernzerhof (PBE) formulation of generalized gradient approximation (GGA)~\cite{ref4}. In addition, for \textit{3d}  Mn orbitals, the Hubbard term correction (GGA+U)~\cite{DFTU}, was adopted. An on-site U parameter of 5 eV, assigned to Mn impurity in Ref.~\citep{ref031}, was considered. The core potential was approximated by the projected augmented wave (PAW) scheme~\cite{ref3}. A cutoff energy of 400 eV for the plane-wave basis set, was found sufficient to achieve a few meV convergence in energy in conjunction with a Brillouin zone sampling of 4$\times$4$\times$1 Gamma-centered Monkhorst-Pack grids.  Finer grids (8$\times$8$\times$1) were used for density of states investigations. The criteria of atom force convergence, used for the structure relaxations, was fixed to 0.02 eV/$\angstrom$.\\

A distance of 20 $\angstrom$ between adjacent MoS$_2$ MLs in perpendicular direction was considered to eliminate spurious interactions resulting from the periodic boundary conditions. Three cases of Mn doping were adopted: an isolated Mn atom per supercell, two Mn atoms per supercell  and three Mn atoms per supercell. In order to significantly reduce long range magnetic interaction between dopants in neighboring cells, a supercell of size 5$\times$5$\times$1 were used to contain one and two Mn dopants while for three Mn dopants, we have considered a 7$\times$7$\times$1 supercell.\\  

The Mn impurities were placed in different positions inside the supercells. The exchange interaction among them was evaluated by the exchange energy, $\Delta E=E_{\textrm{FM}}-E_{\textrm{AFM}}$. $\Delta E$ is the energy difference between the parallel and antiparallel impurity spin orientations. $E_{\textrm{FM}}$ and $E_{\textrm{AFM}}$ are the DFT total energies of self consistent calculations for the FM and AFM configurations, respectively. The magnetic coupling nature (FM or AFM) and its strength was determined by the sign and amount of the exchange energy, respectively. It should be noted that our aim is to evaluate the exchange interaction through $\Delta E$ between a few dopants inside the supercell. A large negative $\Delta E$ indicates a large FM coupling with a relatively high Curie temperature~\cite{ref037,ref038}.

\section{Results and discussion}
\subsection{Single Mn atoms in MoS$_2$ monolayer}
\begin{figure}
  \begin{center}
      \includegraphics[width=0.4\textwidth]{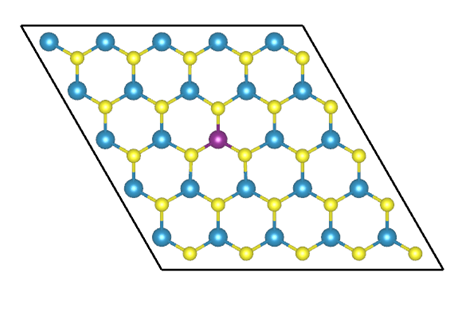} 
  \end{center}
       \caption{ A 4\%-Mn atom doped ML is represented by $5\times 5\times 1$ supercell, the yellow atoms are the Sulfur (S), the blue atoms are the Molybdenum (Mo) and the Manganese atom is denoted by purple color.}
      \label{fig(0)} 
 \end{figure}

We begin our work by examining the electronic properties of an individual Mn dopant in a MoS$_2$ ML at a doping concentration of  4\%, see figure \ref{fig(0)}. In this case, Mn impurities are well separated by a distance equal to 16.25 $\angstrom$. Therefore, one can assume that Mn impurities do not interact with each others.\\
%%%%%%%%%%%%%%%%%%%%%%%%%%%%%%%%%%%%%%%%%%%%%%%%%%%%%%%%%%%%%%%%%%%%%%%%%%%%%%%%%%%%
%few meV
\begin{figure*} 
  \begin{center}
      \subfloat[]{
      \includegraphics[width=0.35\textwidth]{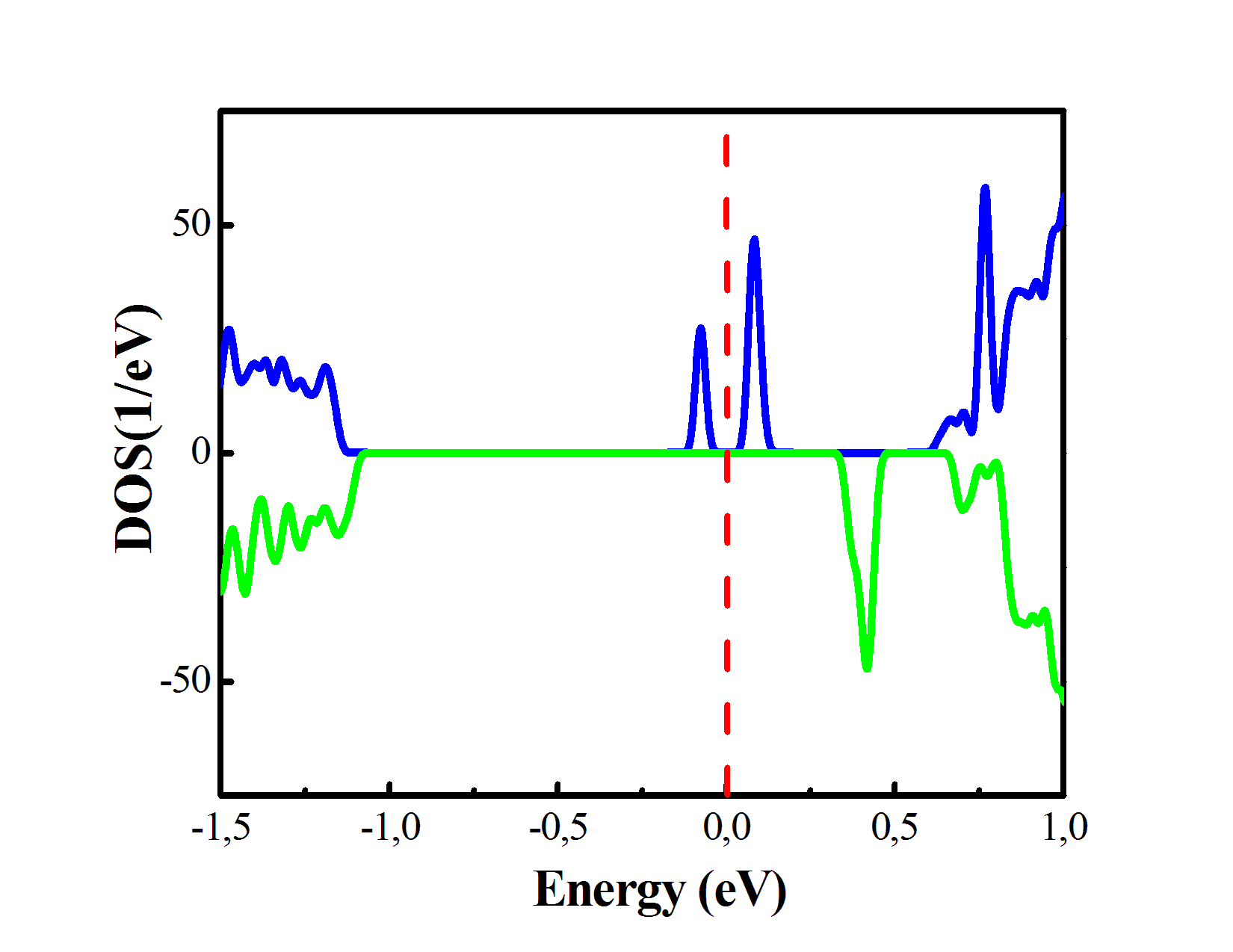}
      \label{fig(11)}
                         }
        \subfloat[]{
      \includegraphics[width=0.35\textwidth]{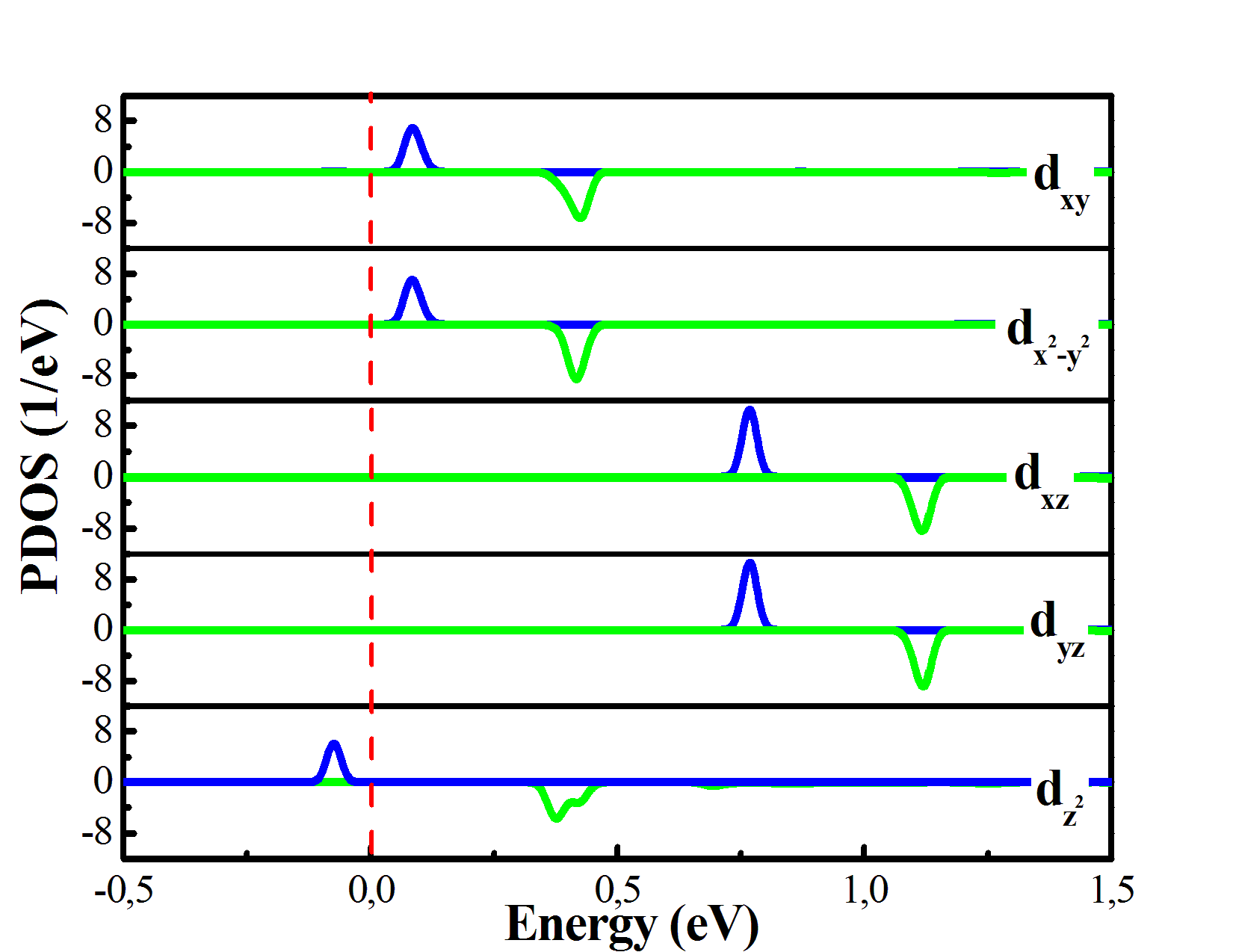}
      \label{fig(12)}
                         } 
    \end{center}
   \captionsetup{format=plain,labelsep=period}                                                      
    \caption{ a) The spin-resolved total electron density of states (DOS) and b) the projected electron density of states (PDOS) of Mn-doped MoS$_{2}$. The blue and green lines denote the spin-up and spin-down channels, respectively. The Fermi level is indicated by the red line.}
    \label{fig:fig(1)}
\end{figure*}
%%%%%%%%%%%%%%%%%%%%%%%%%%%%%%%%%%%%%%%%%%%%%%%%%%%%%%%%%%%%%%%%%%%%%%%%%%%%%%%%%%%%%%%%%%%%%%%
 The splitting of \textit{3d} orbital of Mn impurity in  the MoS$_{2}$ ML are first investigated. The spin-resolved total electron density of states (DOS) and the projected electron density of states (PDOS), are shown in \ref{fig(11)} and \ref{fig(12)} figures, respectively.  As clearly seen in figure \ref{fig(11)}, highly localized states occur in the band gap near the conduction band minimum. The origin of these states is the \textit{3d} Mn orbitals, figure \ref{fig(12)}.  The latter split into three groups the in-plane (\textit{d$_{xy/x^2-y^2}$}) orbitals, the out-of-plane (\textit{d$_{xz/yz}$}) orbitals and the perpendicular \textit{d$_{z^2}$} orbital.  The only occupied states
are the spin up \textit{d$_{z^2}$} orbital which gives rise to a total magnetic moment (MM) of 1$\mu_B$. All these results are in good agreement with previous studies~\cite{ref6,ref7,ref8,ref9}.
\subsection{Double-substitution configurations of Mn atoms in MoS$_2$ monolayer}
\begin{figure*} %[h]
  \begin{center}
    \subfloat[\textbf{-0.18 eV, 5.7 $\angstrom$}]{
      \includegraphics[width=0.4\textwidth]{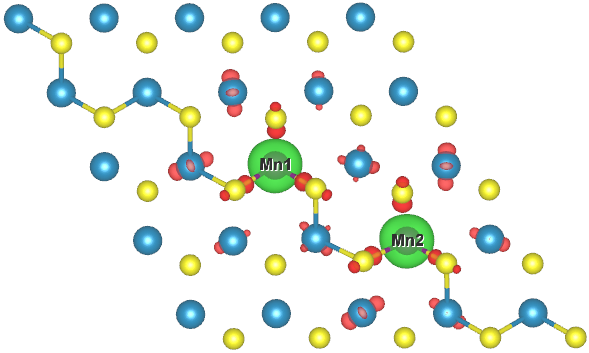}
      \label{fig(21)}
                         }
    \subfloat[\textbf{-0.05 eV, 11.3 $\angstrom$}]{
      \includegraphics[width=0.4\textwidth]{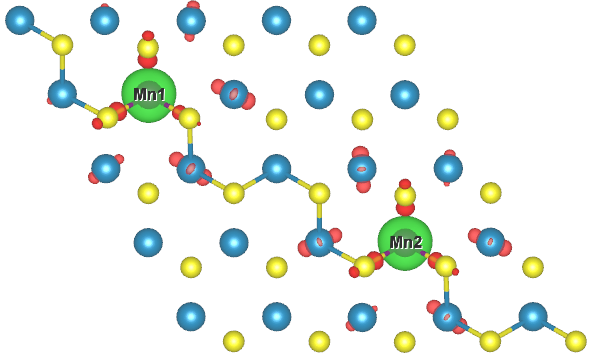}
      \label{fig(22)}
                         }
\qquad     
          \subfloat[\textbf{-0.14 eV, 3.5 $\angstrom$}]{
      \includegraphics[width=0.4\textwidth]{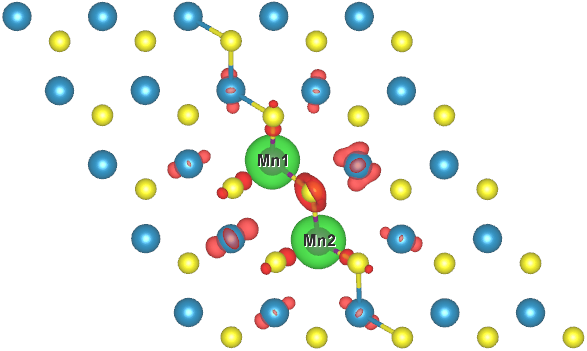}
      \label{fig(23)}
                         }                    
    \subfloat[\textbf{-0.15 eV, 6.5 $\angstrom$}]{
      \includegraphics[width=0.4\textwidth]{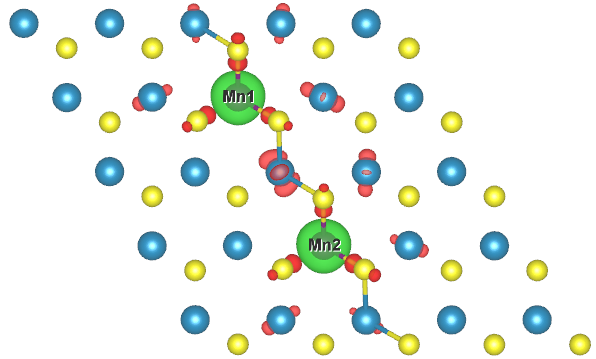}
      \label{fig(24)}
                         }                 
  \end{center}
   \captionsetup{format=plain,labelsep=period} 
      \caption{Spin density isosurface distributions for 8\%-Mn-doped MoS$_2$ in 5$\times$5$\times$1 supercells.  The upper supercells contain each two Mn atoms in two armchair positions while the lower supercells contain each two Mn atoms in two zigzag positions. The energy differences between the AFM and FM coupling states ($\Delta E$) and Mn-Mn distances are listed for each configuration. The green and red isosurfaces represent positive and negative spin density, respectively. The iso-surface value is $0.015$ e$\angstrom^{-3}$.}
      \label{fig(2)}
\end{figure*}

In this section, two Mo atoms, from a 5$\times$5$\times$1 supercell, are replaced by two Mn atoms (noted Mn$_1$ and Mn$_2$ in figure \ref{fig(2)}), which represents 8\% doping concentration in  MoS$_2$ ML. In this case, the available substitution positions suggest four inequivalent configurations as plotted in figure \ref{fig(2)}. We have the following: (i) the \textit{configuration a} (figure \ref{fig(21)}), in which the Mn pairs are placed on 2$^{nd}$ NN  positions of an armchair chain, with a Mn-Mn separation of $5.7$ $\angstrom$; (ii) the \textit{configuration b} (figure \ref{fig(22)}) in which, the Mn pairs are placed on 4$^{nd}$ NN  positions of an armchair chain, with a Mn-Mn separation of $11.3$ $\angstrom$; (iii) the \textit{configuration c} (figure \ref{fig(23)}) in which the two Mn atoms are placed on two consecutive  positions of a zigzag chain where the separation is equal to $3.5$ $\angstrom$; (iv) the \textit{configuration d} (figure \ref{fig(24)}) in which the Mn pairs are placed on 2$^{nd}$ NN substitution positions of a zigzag chain and separated by $6.5$ $\angstrom$.\\
%%%%%%%%%%%%%%%%%%%%%%%%%%%%%%%%%%%%%%%%%%%%%%%%%%%%%%%%%%%%%%%%%%%%%%%%%%%%%%%%%%%%%%%%%%%%%%%%
%%%%%%%%%%%%%%%%%%%%%%%%%%%%%%%%%%%%%%%%%%%%%%%%%%%%%%%%%%%%%%%%%%%%%%%%%%%%%%%%%%%%%%%%%%%%%%%

To get an idea about the stability of different configurations, their relative energies  are listed in Table \ref{tab1}. The lowest energy is corresponding to configuration \textit{c} which contains the closest Mn impurities. This is followed by the configurations \textit{d}, \textit{b} and \textit{a}. One can notice that the configurations with armchair position substitutions are less stable than those with zigzag position substitutions. Furthermore, we find that all double-doping configurations have a common total MM of $\sim$ 2$\mu_B$. This value is similar to that found by many previous reports \citep{ref7,ref8,ref9,ref10}.\\

%%%%%%%%%%%%%%%%%%%%%%%%%%%%%%%%%%%%%%%%%%%%%%%%%%%%%%%
\begin{table*}
\captionsetup{format=plain,labelsep=period}
\caption{ Distance between impurities $L_{\text{Mn}_1-\text{Mn}_2}(\angstrom)$, total MMs, local MMs (for Mn$_1$ and Mn$_2$) and magnetic energies of various configurations which are shown in figure \ref{fig(2)}.}
  \centering
  \resizebox{1.\textwidth}{!}{
  \begin{tabular}{cccccccccccccc}
  \hline
  \hline
  Double doping-configurations&& Separation distances && Relative energy& &Total magnetic && Local magnetic  & moment ($\mu_B$)  &&  $\Delta E$\\
   && $L_{\text{Mn}_1-\text{Mn}_2}(\angstrom)$  && (eV) && moment ($\mu_B$) && Mn$_1$ & Mn$_2$ && (eV) \\
 \hline
  a && 5.7 &&0.51 &&2.00  && 3.18   &3.21    &&-0.18\\ 

  b && 11.3 &&0.51 &&1.99    && 3.19    & 3.21  &&-0.05\\  

  c && 3.5 &&0.00&&2.00  &&3.22  &3.22    &&-0.14\\

  d && 6.5  && 0.40&& 2.00 &&3.20    & 3.20   &&-0.15\\
   \hline
   \hline
\end{tabular}
}
\label{tab1}
\end{table*}
%%%%%%%%%%%%%%%%%%%%%%%%%%%%%%%%%%%%%%%%%%%%%%%%%%%%%%%%%

In order to figure out the magnetic coupling nature in each doping configuration, their exchange energies are presented in table \ref{tab1}. The exchange energy is found negative for the four configurations which means that they have stable FM states. In particular, for Mn-Mn separations equal or less than $5.7$ $\angstrom$, the double-doping configurations have large $\Delta E$, above 0.10 eV, which should stabilize their FM nature at high temperature. The FM exchange interactions depend on the doping configuration as well as Mn-Mn separations. More specifically, the configuration \textit{a} shows the strongest FM exchange coupling even greater than configuration \textit{c} which has the closest Mn pair. This dominance  of the FM interaction between Mn impurities in the configuration a is also found in Ref.~\cite{ref10}.\\

 Using the SSP model, we investigate the magnetic coupling of two Mn dopants in different positions. To this end, the spin-polarized charge density isosurface distributions of the FM state for the different cases are shown the figure \ref{fig(2)}. According to the SSP model the Mn-Mn FM coupling is based on the interaction with the spin-polarized neighboring atoms, namely S and Mo atoms~\cite{ref032}. The Mn-Mn FM coupling is justified by the fact that the two dopants are identical and both induce the same type of polarization on nearby atoms~\cite{ref032}. For all doping configurations, the induced spins on the nearby host atoms are antiparallel to those of the Mn dopants. The same spin density behavior has been found in previous  studies~\cite{ref7,ref8,ref9,ref10}. Furthermore, as shown in the table \ref{tab1}, the local MMs of the dopants Mn are larger than the total MM of the doped ML. In fact, the AFM coupling between the impurities of Mn and their host NN atoms is at the origin of the reduction of the total MM. The latter result is in agreement with Ref.~\cite{ref10}.\\

 As we mentioned before, the FM coupling  depends on the strength of the induced polarization on the NN anions mediating the two dopants. Indeed, the strength of the induced antiparallel spin density that resides on the Mn-Mn mediating S and Mo, differs from one configuration to another, see figure \ref{fig(2)}. For the different configurations in figure \ref{fig(2)}, we classify the induced polarization between Mn dopants  from the most important to the weakest as \textit{c} than \textit{d}, \textit{a} and \textit{b} configurations. For configuration \textit{b} (figure \ref{fig(22)}), the distance between two Mn dopants is $11.3$ $\angstrom$ which is so large that the SSP processes between them is weak and therefore one can consider them to be almost isolated.  For the rest of cases, we notice that the FM coupling is inversely proportional to the spin density magnitude that mediates the two Mn impurities. In particular, for the case of configuration \textit{c} (figure \ref{fig(23)}) there is a strong AFM coupling between  Mn and the mediated S and Mo atoms, which weakens the FM interaction between the two dopants. For configuration \textit{d}  (figure \ref{fig(24)}), Mn1 and the mediated host material atoms are less AFM coupled compared to  configuration \textit{c}. Thus, configuration \textit{d} shows more stable Mn1-Mn2 FM coupling than \textit{c} configuration. Unlike the latter two configurations \textit{c} and \textit{d}, \textit{a} has the lowest AFM coupling between the dopants and the mediated atoms, which promotes its FM stability. It can be said that the antiparallel spin density in the middle of the dopants filters the FM interaction between the two dopants, more than it is important, more than the FM exchange is weak. In general, it can be said that the Mn-Mn mediating antiparallel spin density screens the FM interaction between the two dopants, more than it is important, more than the Mn-Mn FM exchange is weak.\\
\subsection{Triple-substitution configurations of Mn atoms in MoS$_2$ monolayer}
\begin{figure} %[h]
  \begin{center}
    \subfloat[]{
      \includegraphics[width=0.09\textwidth]{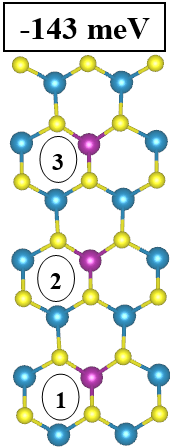}
      \label{fig(51)}
                         }
    \subfloat[]{
      \includegraphics[width=0.09\textwidth]{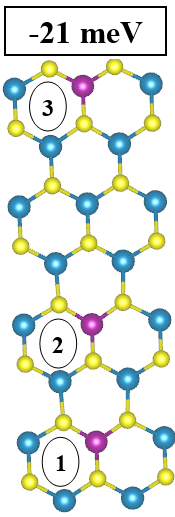}
      \label{fig(52)}
                         }    
          \subfloat[]{
      \includegraphics[width=0.09\textwidth]{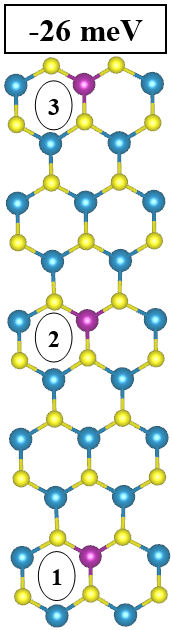}
      \label{fig(53)}
                         }
\qquad                    
    \subfloat[]{
      \includegraphics[width=0.09\textwidth]{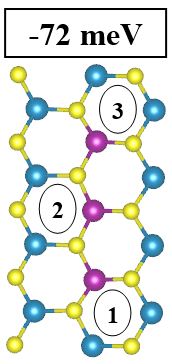}
      \label{fig(54)}
                         }
    \subfloat[]{
      \includegraphics[width=0.09\textwidth]{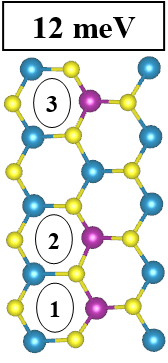}
      \label{fig(55)}
                         }
   \subfloat[]{
      \includegraphics[width=0.09\textwidth]{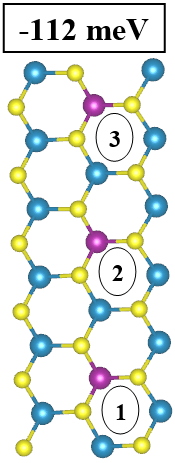}
      \label{fig(56)}
                         }  
\qquad                          
          \subfloat[]{
      \includegraphics[width=0.12\textwidth]{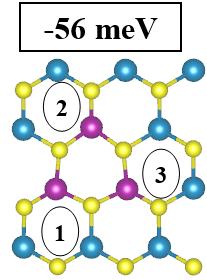}
      \label{fig(57)}
                         }                     
    \subfloat[]{
      \includegraphics[width=0.12\textwidth]{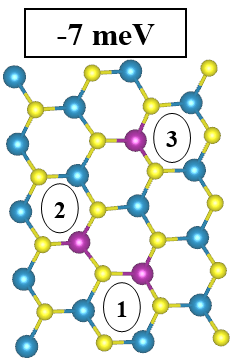}
      \label{fig(58)}
                         }   
    \subfloat[]{
      \includegraphics[width=0.12\textwidth]{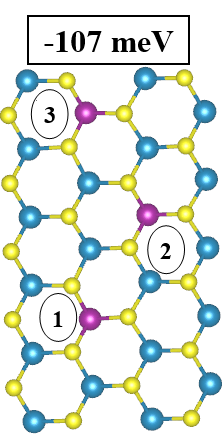}
      \label{fig(59)}
                         }
\qquad                          
          \subfloat[]{
      \includegraphics[width=0.12\textwidth]{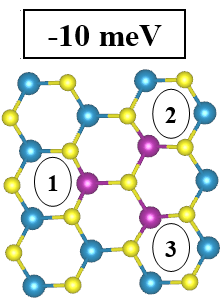}
      \label{fig(501)}
                         }                     
    \subfloat[]{
      \includegraphics[width=0.12\textwidth]{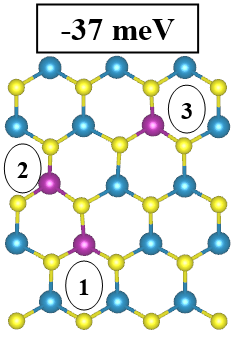}
      \label{fig(502)}
                         }   
    \subfloat[]{
      \includegraphics[width=0.12\textwidth]{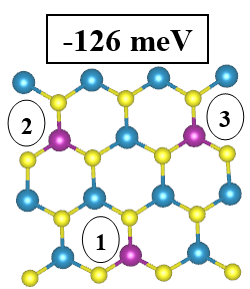}
      \label{fig(503)}
                         }   
  \end{center}                                                                                      
\caption{Schematics showing various configurations of triple-dopant configurations (TDCs) that were considered. Only the relevant portions of the 7$\times$7$\times$1 supercells are shown, for clarity. Blue, yellow, and purple balls represent Mo, S, and Mn atoms, respectively. In each case, the Mn atoms are denoted by $(1)$, $(2)$ and $(3)$. The energy differences between the FM state and the energetically-closest AFM state ($\Delta E^*$) are listed for each doping configuration. }
    \label{fig(5)}
\end{figure}

In this section, we study the magnetic properties of three substitutional  Mn impurities that replace three close Mo atoms of MoS$_2$ ML. Unlike the previous section, we expand the supercell to 7$\times$7$\times$1, which means a $6.12\%$ doping concentration. A large number of configurations are considered in which the Mn atoms are  placed in various relative Mo-sites. In figure \ref{fig(5)}, we summarize the resulting triple-doping configurations (TDCs). In each case, for clarity, we show the relevant portion of the 7$\times$7$\times$1 supercell that contains the three dopants (figure \ref{fig(5)}).\\

 Beginning by treating the TDCs stabilities, their relative energies are listed in table \ref{tab3}. Our calculations indicate that the configurations where the Mn impurities are placed at NN positions, namely configuration \textit{d}, \textit{g} and \textit{j}, are more energetically favored compared to the rest of configurations. In other words, the three Mn dopants prefer to stay close to each other. In particular, the TDC lowest energy is associated to the configuration \textit{j}, in which the impurities are bonded to the same S atom, see figure \ref{fig(501)}. These results suggest that high concentration doping can lead to the clustering of Mn impurities in the MoS$_2$ ML.\\
%%%%%%%%%%%%%%%%%%%%%%%%%%%%%%%%%%%%%%%%%%%%%%%%%%%%%%%%%%%%%%%%%%%%%%%%%%%%%%%%%%%%%%%%%%
%%%%%%%%%%%%%%%%%%%%%%%%%%%%%%%%%%%%%%%%%%%%%%%%%%%%%%%%%%%%%%%%%%%%%%%%%%%%%%%%%%%%%%%%
\begin{figure*} %[h]
  \begin{center}
    \subfloat[]{
      \includegraphics[width=0.5\textwidth]{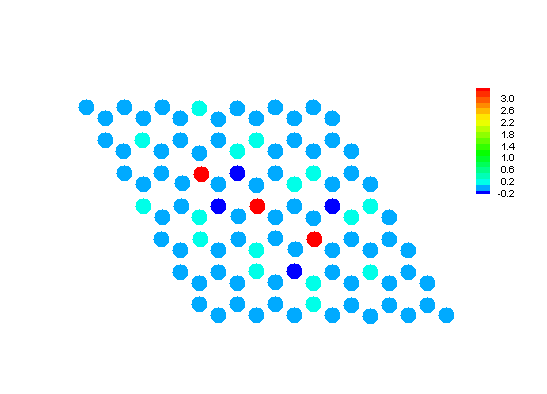}
      \label{fig(61)}
                         }
    \subfloat[]{
      \includegraphics[width=0.5\textwidth]{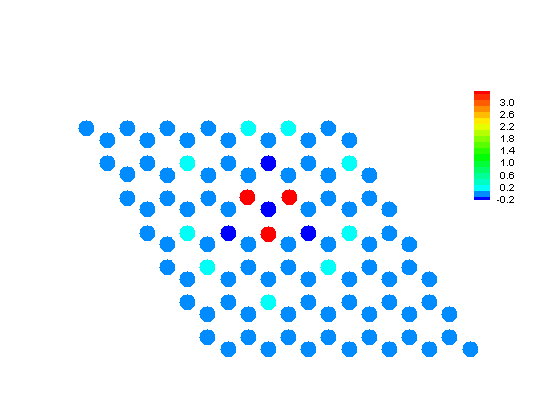}
      \label{fig(62)}
                         }
\qquad 
     \subfloat[]{
      \includegraphics[width=0.5\textwidth]{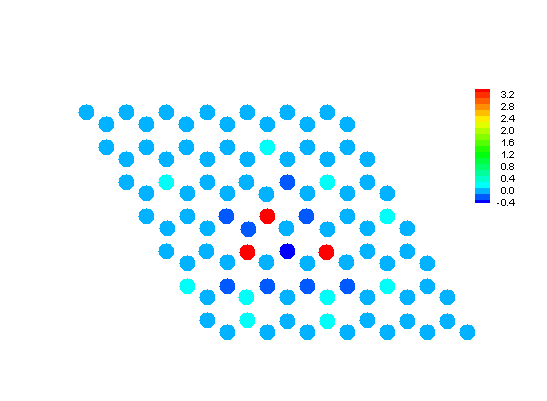}
      \label{fig(63)}
                         }
 \end{center}
 \captionsetup{format=plain,labelsep=period}                                                                                         
\caption{ The local  spin magnetic moments in Bohr magneton of Mn impurities and of their surrounding neighboring atoms as obtained from the FM state for a) configuration \textit{a}, b) configuration \textit{j}, and c) configuration \textit{h} depicted in Fig.~\ref{fig(5)}.}
    \label{fig(6)}
\end{figure*}
%%%%%%%%%%%%%%%%%%%%%%%%%%%%%%%%%%%%%%%%%%%%%%%%%%%%%%%%%%%%%%%%%%%%%%%%%%%%%%%%%%%%%%%%%%%%%%% 
\begin{figure} 
  \begin{center}
  \includegraphics[width=0.5\textwidth]{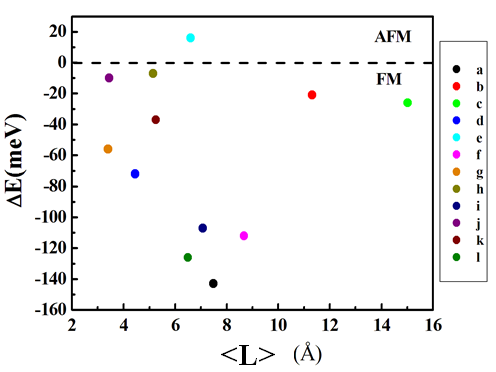}
  \end{center}
      \label{fig(80)}
 \caption{a) The gain in energy of the FM ordering over AFM ordering of spins for the three Mn impurities as a function of the average separation distance, $<\mathrm{L}>$, between the impurities. the negative (positive) energy corresponds to FM (AFM) ordering being more stable.}                        
     \label{fig(8)}                
\end{figure}
%%%%%%%%%%%%%%%%%%%%%%%%%%%%%%%%%%%%%%%%%%%%%%%%%%%%%%%%%%%%%%%%%%%%%%%%%%%%%%%%%%%%%%%%%%%%%%%%
 It should be noted also, as in table \ref{tab3}, that while the NN TDCs, \textit{d} \textit{g} and \textit{j}, have a large total MMs of $\sim$ 5 $\mu_B$, the rest of configurations have a total MMs of just 3 $\mu_B$. This difference of total MM originates from the different environments of dopants. The total MMs result from the competition between the Mn positive local MMs and the negative local MMs of the host atoms. For instance, local  spin magnetic moments of Mn impurities and their surrounding atoms for configurations \textit{a}, \textit{j} and  \textit{h} are depicted in figure \ref{fig(6)}. For the NN TDCs, local MMs of $\sim$ 10 $\mu_B$ and $\sim$ -3.8 $\mu_B$ were found for the Mn impurities and the host atoms, respectively. However, for the rest of configurations, local MMs of $\sim$ 9.7 $\mu_B$ and $\sim$ -5 $\mu_B$ were found for the Mn impurities and the host atoms, respectively. The reduced MM of the environment of dopants in case of NN TDCs causes the increasing of their overall MMs.\\

 The energy of the different considered magnetic states is denoted $E(\sigma_1,\sigma_2,\sigma_3)$, where $\sigma_i$ represents the spin orientation of Mn$_i$ impurities.  Obviously, $E(\uparrow,\uparrow,\uparrow)$ is the energy of the FM state. The enumeration of Mn$_i$ is shown in figure \ref{fig(5)} for all TDCs. For the sake of comparison, we denote the closest Mn neighbors Mn$_1$ and Mn$_2$.\\
  %%%%%%%%%%%%%%%%%%%%%%%%%%%%%%%%%%%%%%%%%%%%%%%%%%%%%%%%%%%%%%%%%%%%%%%%%%%%%%%%%%%%%%%%%%%
 \begin{table*}
\captionsetup{format=plain,labelsep=period} 
\caption{Relative energies, total magnetic moments (MMs) and the energy differences between FM and non-FM spin configurations, $\Delta E$, of various triple doping configurations. The asterisk indicates the energetically closest spin configuration to the ground state.}
  \centering
  \resizebox{0.7\textwidth}{!}{
  \begin{tabular}{ccccccccccccccccccc}
  \hline
  \hline
  Configurations & Relative energy (eV)& MM($\mu_B$)& Spin configurations & $\Delta E$ (eV) \\
 \hline
  a  &2.11 &2.99 &$E(\uparrow,\uparrow,\uparrow)-E(\uparrow,\downarrow,\uparrow)$  & -0.24 \\
  & & &$E(\uparrow,\uparrow,\uparrow)-E(\uparrow,\uparrow,\downarrow)^*$  & -0.14 \\ 
  b &2.10 &2.99 &$E(\uparrow,\uparrow,\uparrow)-E(\uparrow,\downarrow,\uparrow)$  & -0.13\\
   & & &$E(\uparrow,\uparrow,\uparrow)-E(\downarrow,\uparrow,\uparrow)$  & -0.12 \\
  & & &$E(\uparrow,\uparrow,\uparrow)-E(\uparrow,\uparrow,\downarrow)^*$  & -0.02 \\ 
  c &2.26 &2.99&$E(\uparrow,\uparrow,\uparrow)-E(\downarrow,\uparrow,\uparrow)$  & -0.04 \\ 
    & & &$E(\uparrow,\uparrow,\uparrow)-E(\uparrow,\downarrow,\uparrow)^*$  & -0.02  \\
  d &0.71&5.00&$E(\uparrow,\uparrow,\uparrow)-E(\uparrow,\downarrow,\uparrow)$  & -0.14 \\ 
    & & &$E(\uparrow,\uparrow,\uparrow)-E(\uparrow,\uparrow,\downarrow)^*$  & -0.07 \\
  e &2.45&1.00&$E(\uparrow,\uparrow,\uparrow)-E(\uparrow,\downarrow,\uparrow)$ & -0.07 \\ 
   & & &$E(\uparrow,\uparrow,\uparrow)-E(\uparrow,\downarrow,\uparrow)$  & -0.07 \\
    & & &$E(\uparrow,\uparrow,\uparrow)-E(\uparrow,\uparrow,\downarrow)^*$  & 0.01 \\ 
  f &1.88 &2.99&$E(\uparrow,\uparrow,\uparrow)-E(\uparrow,\downarrow,\uparrow)$  & -0.25 \\
  & & &$E(\uparrow,\uparrow,\uparrow)-E(\uparrow,\uparrow,\downarrow)^*$  & -0.11 &\\  
  g &0.45&4.97&$E(\uparrow,\uparrow,\uparrow)-E(\uparrow,\downarrow,\uparrow)^*$  & -0.05 \\ 
  h &1.30&3.00 &$E(\uparrow,\uparrow,\uparrow)-E(\uparrow,\downarrow,\uparrow)$  & -0.14\\
    & & &$E(\uparrow,\uparrow,\uparrow)-E(\downarrow,\uparrow,\uparrow)$  & -0.09\\
  & & &$E(\uparrow,\uparrow,\uparrow)-E(\uparrow,\uparrow,\downarrow)^*$  & -0.00 \\   
  i  &2.90&3.00&$E(\uparrow,\uparrow,\uparrow)-E(\uparrow,\downarrow,\uparrow)$  & -0.22\\ 
  & & &$E(\uparrow,\uparrow,\uparrow)-E(\uparrow,\uparrow,\downarrow)^*$  & -0.10 \\ 
  j &0.00&5.00&$E(\uparrow,\uparrow,\uparrow)-E(\uparrow,\downarrow,\uparrow)^*$  & -0.01 \\ 
  k &1.34&4.99&$E(\uparrow,\uparrow,\uparrow)-E(\uparrow,\downarrow,\uparrow)$  & -0.10 \\
  & & &$E(\uparrow,\uparrow,\uparrow)-E(\downarrow,\uparrow,\uparrow)$  & -0.11 \\ 
    & & &$E(\uparrow,\uparrow,\uparrow)-E(\uparrow,\uparrow,\downarrow)^*$  & -0.03 \\  
  l &1.83&3.00&$E(\uparrow,\uparrow,\uparrow)-E(\uparrow,\uparrow,\downarrow)^*$  &-0.12 \\ 
   \hline
   \hline
\end{tabular}
}
\label{tab3}
\end{table*}
%%%%%%%%%%%%%%%%%%%%%%%%%%%%%%%%%%%%%%%%%%%%%%%%%%%%%%%%%%%%%%%%%%%%%%%%%%%%%%%%%%%%%%%%%
 
  In Table \ref{tab3}, we present the energy differences, $\Delta E$, between FM and non-FM configurations for the various TDCs. In each TDC, the energy difference between the ground state and the energetically-closest magnetic state is asterisked $\Delta E^*$ in table \ref{tab3}. The FM state is found to be stable for all TDCs except configuration \textit{e} which has an AFM ground state. Furthermore, for each TDC the energy differences, $\Delta E$, where the AFM energies have the spin configurations ($\downarrow,\uparrow,\uparrow$) or ($\uparrow,\downarrow,\uparrow$), are the most important. This result proves that the NN impurities, Mn$_1$ and Mn$_2$ as shown in figure \ref{fig(5)}, have a strong FM coupling. However, the presence of a third close Mn impurity tends in general to destabilise the FM state. For configuration \textit{e}, the ground state becomes even AFM ($\uparrow,\uparrow,\downarrow$). Clearly, modifying the doping configurations alters the stability of the FM state and with that certainly the strength and nature of the inter-impurities magnetic interactions.\\
  
  In particular, depending on the FM stability and the geometric similarities, three TDC groups stand out. (i) The first TDC group is formed by configurations \textit{a}, \textit{f}, \textit{i}, and \textit{l}.  Here, in each configuration, at least one Mn dopant is placed at the mediator of the segment formed by the other two Mn dopants. At least two pairs of Mn atoms are 2$^{nd}$ NNs, see figures \ref{fig(51)}, \ref{fig(56)}, \ref{fig(59)}, \ref{fig(503)}. The TDCs in this group have the most stable FM states compared to the rest of configurations. Indeed,  $\Delta E^*$ are of the order of hundreds meV which is comparable to the \textit{a}, \textit{c}, and \textit{d} double-doping configuration energies. Similar to the case of double-doping configurations, Mn dopants on the 2$^{nd}$ NN positions of an armchair chain, configuration \textit{a}, exhibits the most stable FM-state. Ordering the Mn dopants on the MoS$_2$ ML in this particular set of configurations can increase the temperature stability of the FM state.
(ii) The second group includes configurations \textit{d}, \textit{g} and \textit{j}. In contrast to the first group,  here the Mn atoms are placed in the NN positions (see figures \ref{fig(54)}, \ref{fig(57)}, \ref{fig(501)}). In this case, the FM interaction of the TDC is weak since $\Delta E^*$ is of the order of tens meV.  In this group, the lowest  $\Delta E^*$ is found for configuration \textit{j}. The origin of the reduction of the overall FM state of configuration \textit{j} is attributed to the strong AFM coupling of the three dopants with the mediating atoms, see figure \ref{fig(62)}. Therefore, clustering of Mn impurities in the ML is not preferable if we want to get strong ferromagnetism. This means that although clustering is energetically favorable, we need to avoid it. This result is consistent with an observation of Jieqiong \textit{et al.}~\cite{ref14} in which those Mn with NN form Mn clusters are typically AFM.
(iii) A weak FM interaction is also found in the third group which includes the configurations \textit{b}, \textit{e}, \textit{h}, and \textit{k}. Here, the three dopants are placed at different distances from each others (see figures \ref{fig(52)}, \ref{fig(55)}, \ref{fig(58)}, \ref{fig(502)}). In other words, one Mn dopant is far from the other two Mn which are close to each other. As shown in figure \ref{fig(63)} for configuration \textit{h}, a strong negative local MM of $\sim$ $-0.4$ $\mu_B$ resides on the mediating Mo between the two close Mn and the far Mn dopant. This explains why  the third far Mn reduces the FM state of configuration \textit{i}. In Ref.~\cite{ref034}, according to SSP, if one dopes with two different magnetic impurities, the spin polarization on the mediating host atom will take the characteristics of the strongest polarization induced on it by the two neighboring impurities. In our case, the situation is quite similar, the two close  Mn atoms act as one atom that induces a strong AFM polarization on the host atoms which dictates the spin polarization of the third Mn atom. This favors the tendency towards a weak FM coupling or even for an AFM coupling, as obtained for configuration \textit{e}, between the two close Mn impurities and the far Mn impurity.\\
     
To get a better insight on  the ferromagnetic stability of the TDCs, we plot in \ref{fig(8)}, $\Delta E^*$ as a function of the average separation distances between the impurities. One notices that the most stable FM state is realized for the average separation distance ranging from 6 to 9 $\angstrom$. Outside these inter-impurity distances, the FM state is weakened but maintained up to large distances.  Our findings are in agreement with the experiment reported in Ref.~\cite{ref14}, which indicates that only Mn impurities that are at suitable distances can order ferromagnetically. Figure \ref{fig(8)} shows furthermore that the doping configuration plays a criticial role in the magnetic stability of the impurities complexes since for comparable averaged inter-impurity distances the energy differences can be very different. For instance, $\Delta E^*$ for configuration \textit{j}   (-0.01 eV)  is about 5 times smaller than that of configuration \textit{g} (-0.05 eV) although they are both characterized by the same averaged Mn-Mn distance.\\ 
  
After our discussion of the magnetic stability of the various complexes, we complete our study by evaluating the magnetic interactions among the Mn atoms. To this end, we map the energy differences obtained from first-principles for the various studied magnetic states  to those of the classical Heiseberg model,  $H=-\frac{1}{2}\sum_{i\neq j}J_{ij}\bm{e}_i\bm{e}_j$. Here $\bm{e}_i$ is the unit vector defining the direction of Mn atomic  MM at site $i$ and $J_{ij}$ are the magnetic exchange coupling constants between the local moments at Mn-sites $i$ and $j$.  For the double-doping configuration, $J_{12}^{double} =\displaystyle -\frac{\Delta E}{2}$. For the triple-doping configurations, we have different cases. When Mn-complex form an equilateral triangle, the three possible AFM states are degenerate, i.e.  $(\downarrow,\uparrow,\uparrow)\equiv(\uparrow,\downarrow,\uparrow)\equiv(\uparrow,\uparrow,\downarrow)$ and $J_{12}^{triple}=J_{23}^{triple}=J_{13}^{triple}=\displaystyle -\frac{\Delta E}{4}$. In the case of two inequivalent AFM states,  $J_{12}^{triple}=J_{23}^{triple}=\displaystyle-\frac{\Delta E_1}{4}$ and $J_{13}^{triple}=\displaystyle\frac{\Delta E_1-2\Delta E_2}{4}$ where $\Delta E_1=E(\uparrow,\uparrow,\uparrow)-E(\uparrow,\downarrow,\uparrow)$ and $\Delta E_2=E(\uparrow,\uparrow,\uparrow)-E(\uparrow,\uparrow,\downarrow)$.  Finally, when the three AFM states are inequivalent, $J_{12}^{triple}=\displaystyle\frac{-\Delta E_1-\Delta E_2+\Delta E_3}{4}$, $J_{23}^{triple}=\displaystyle\frac{\Delta E_1-\Delta E_2-\Delta E_3}{4}$ and $J_{13}^{triple}=\displaystyle\frac{-\Delta E_1+\Delta E_2-\Delta E_3}{4}$ where $\Delta E_3=E(\uparrow,\uparrow,\uparrow)-E(\downarrow,\uparrow,\uparrow)$.\\	
	
 The estimated values of $J_{ij}$ are listed in table \ref{tab5}. Comparing the constants $J_{ij}^{triple}$ to $J_{ij}^{double}$, we see that some of them remain almost unaltered and some others change significantly.  For instance, the first $J_{ij}$ line in table \ref{tab5} shows that except for the doping configurations \textit{h} and \textit{k}, the exchange coupling constants $J_{ij}^{double}$ of Mn pairs placed on the 2$^{nd}$ NN Mo-site, are comparable to  $J_{ij}^{double}$. Furthermore, for the other two lines, it is clear that there is a large fluctuation of $J_{ij}^{triple}$ with respect to $J_{ij}^{double}$. This reveals the dependence of pairwise magnetic coupling on the pair environment which contains a third dopant as expected from our previous discussion. Interestingly, we note one case where the magnetic interaction becomes AFM: configuration \textit{h} while for configurations \textit{d}, \textit{k} and \textit{e} the magnetic interactions between the furthest apart Mn atoms is negligible.\\
%%%%%%%%%%%%%%%%%%%%%%%%%%%%%%%%%%%%%%%%%%%%%%%%%%%%%%%
\begin{table*}
\captionsetup{format=plain,labelsep=period}
\caption{ Estimated magnetic exchange coupling constants $J_{ij}$ of triple- and double-doping configurations. Each $J_{ij}$ row corresponds to the Mn-pairs with common separation distance L$_{Mn_i-Mn_j}$.}
  \centering
  \resizebox{1.02\textwidth}{!}{
  \begin{tabular}{cccccccccccc}
  \hline
  \hline
  &&&&&&&\\
  & L$_{Mn_i-Mn_j}$ ($\angstrom$) &   \multicolumn{6}{c}{Triple doping-configurations }    & Double doping-configurations \\
  & & \textit{a} & \textit{b}& \textit{h} &\textit{i} & \textit{k}& &\textit{a}\\
 \hline
  &&&&&&&\\
  $J_{ij}$(eV) & $\sim$ 5.7 &\hspace{0.2cm} $J_{12(23)}=$ 0.06 & \hspace{0.2cm} $J_{12}=$ 0.05  & \hspace{0.2cm} $J_{23}=$ 0.01  & \hspace{0.2cm} $J_{12(23)}=$ 0.05   &\hspace{0.2cm}$J_{23}=$ 0.00 &\hspace{0.2cm}& \hspace{0.2cm}$J_{12}=$ 0.09\\ 
  &&&&&&&\\
  \hline
  \hline
  &&&&&&&\\
 & & \textit{d} & \textit{e}& \textit{g} &\textit{j} & \textit{h}& \textit{k}&\textit{c}\\
  \hline
  &&&&&&&\\
 $J_{ij}$(eV)& $\sim$ 3.5 & $J_{12(23)}=$ 0.03 &$J_{12}=$ 0.04  &$J_{12(23,13)}=$ 0.01  & $J_{12(23,13)}=$ 0.02   &$J_{12}=$ 0.06    &$J_{12}=$ 0.04 &$J_{12}=$ 0.07\\   
 &&&&&&&\\
  \hline
  \hline
  &&&&&&&\\
   & & \textit{d} & \textit{e}& \textit{f} &\textit{h} & \textit{k}& \textit{l}&\textit{d}\\
  \hline
  &&&&&&&\\
 $J_{ij}$(eV)& $\sim$ 6.5 & $J_{13}=$ 0.00 & $J_{23}=$ -0.00  &$J_{12(23)}=$ 0.06  & $J_{13}=$ -0.01   &$J_{13}=$ 0.01    &$J_{12(23,13)}=$ 0.03&$J_{12}=$ 0.07\\ 
 &&&&&&&\\
   \hline
   \hline
\end{tabular}
}
\label{tab5}
\end{table*}
%%%%%%%%%%%%%%%%%%%%%%%%%%%%%%%%%%%%%%%%%%%%%%%%%%%%%%%%%
 
 \section{Conclusion}
Performing DFT+U calculations, we show that the magnetic stability and the magnetic exchange interaction between neighboring dopants are very sensitive to the doping-configuration geometry and the dopant separation distances. Our calculations suggest on the one hand that placing Mn dopants at equidistant Mo-sites where the average dopants separation is 6-9 $\angstrom$, enhances the ferromagnetism of Mn-doped MoS$_2$ ML. The Mn impurities that are placed on the 2$^{nd}$ NN Mo-sites of an armchair chain have the strongest FM coupling. On the other hand, the FM exchange interaction is found to be reduced dramatically when we have a Mn impurity close to a Mn-cluster. Interestingly, the ferromagnetic interactions are in general finite for large inter-impurity distances. In addition, the Mn impurities in the closest Mo-sites, clusters, show weak FM coupling. The diversity in the  FM coupling strength for the various doping configuration is due to the strength of antiparallel spin polarized Mo and S atoms that are mediating  the interactions among Mn impurities. When the Mn impurities approach each other the anti-parallel mediating MMs increase, which reduces the FM exchange interaction. It should be noted that, the doping configuration in which the FM exchange is low are found energetically favorable indicating that Mn impurities have the tendency to clustering within the MoS$_2$ ML.  Our results show that doping control is very necessary to take advantage of magnetic properties of this material. This is achievable with atomic manipulation using scanning tunneling microscopy, which in its spin-polarized version allows even to extract magnetic exchange interactions at the atomic scale~\cite{lounis1,lounis2}. This offers the possibility of confirming our predictions.\\
%%%%%%%%%%%%%%%%%%%%%%%%%%%%%%%%%%%%%%%%%%%%%%%%%%%%%%%%%%%%%%%%%%%%%%%%%%%%%%%%%%%%
\section*{ACKNOWLEDGMENT}
I.C.G. thanks the CALMIP initiative for the generous allocation of computational time through project p0812, as well as GENCI-CINES and GENCI-IDRIS for Grant No. 2018-A004096649.\\
S.L. acknowledges funding from the European Research Council (ERC) under the European Union’s Horizon 2020 research and innovation programme (ERC-consolidator Grant No. 681405 DYNASORE).
%%%%%%%%%%%%%%%%%%%%%%%%%%%%%%%%%%%%%%%%%%%%%%%%%%%%%%%%%%%%%%%%%%%%%%%%%%%%%%%%%%%%
\bibliographystyle{apsrev}
\bibliography{references}

\begin{thebibliography}{52}
\expandafter\ifx\csname natexlab\endcsname\relax\def\natexlab#1{#1}\fi
\expandafter\ifx\csname bibnamefont\endcsname\relax
  \def\bibnamefont#1{#1}\fi
\expandafter\ifx\csname bibfnamefont\endcsname\relax
  \def\bibfnamefont#1{#1}\fi
\expandafter\ifx\csname citenamefont\endcsname\relax
  \def\citenamefont#1{#1}\fi
\expandafter\ifx\csname url\endcsname\relax
  \def\url#1{\texttt{#1}}\fi
\expandafter\ifx\csname urlprefix\endcsname\relax\def\urlprefix{URL }\fi
\providecommand{\bibinfo}[2]{#2}
\providecommand{\eprint}[2][]{\url{#2}}

\bibitem[{\citenamefont{Najmaei et~al.}(2013)\citenamefont{Najmaei, Liu, Zhou,
  Zou, Shi, Lei, Yakobson, Idrobo, Ajayan, and Lou}}]{ref01}
\bibinfo{author}{\bibfnamefont{S.}~\bibnamefont{Najmaei}},
  \bibinfo{author}{\bibfnamefont{Z.}~\bibnamefont{Liu}},
  \bibinfo{author}{\bibfnamefont{W.}~\bibnamefont{Zhou}},
  \bibinfo{author}{\bibfnamefont{X.}~\bibnamefont{Zou}},
  \bibinfo{author}{\bibfnamefont{G.}~\bibnamefont{Shi}},
  \bibinfo{author}{\bibfnamefont{S.}~\bibnamefont{Lei}},
  \bibinfo{author}{\bibfnamefont{B.~I.} \bibnamefont{Yakobson}},
  \bibinfo{author}{\bibfnamefont{J.-C.} \bibnamefont{Idrobo}},
  \bibinfo{author}{\bibfnamefont{P.~M.} \bibnamefont{Ajayan}},
  \bibnamefont{and} \bibinfo{author}{\bibfnamefont{J.}~\bibnamefont{Lou}},
  \bibinfo{journal}{Nature materials} \textbf{\bibinfo{volume}{12}},
  \bibinfo{pages}{754} (\bibinfo{year}{2013}).

\bibitem[{\citenamefont{Tan and Zhang}(2015)}]{ref02}
\bibinfo{author}{\bibfnamefont{C.}~\bibnamefont{Tan}} \bibnamefont{and}
  \bibinfo{author}{\bibfnamefont{H.}~\bibnamefont{Zhang}},
  \bibinfo{journal}{Chem. Soc. Rev.} \textbf{\bibinfo{volume}{44}},
  \bibinfo{pages}{2713} (\bibinfo{year}{2015}).

\bibitem[{\citenamefont{Li et~al.}(2016)\citenamefont{Li, Huang, Zhong, Li,
  Wang, Li, and Wei}}]{ref03}
\bibinfo{author}{\bibfnamefont{B.}~\bibnamefont{Li}},
  \bibinfo{author}{\bibfnamefont{L.}~\bibnamefont{Huang}},
  \bibinfo{author}{\bibfnamefont{M.}~\bibnamefont{Zhong}},
  \bibinfo{author}{\bibfnamefont{Y.}~\bibnamefont{Li}},
  \bibinfo{author}{\bibfnamefont{Y.}~\bibnamefont{Wang}},
  \bibinfo{author}{\bibfnamefont{J.}~\bibnamefont{Li}}, \bibnamefont{and}
  \bibinfo{author}{\bibfnamefont{Z.}~\bibnamefont{Wei}},
  \bibinfo{journal}{Advanced Electronic Materials}
  \textbf{\bibinfo{volume}{2}}, \bibinfo{pages}{1600298}
  (\bibinfo{year}{2016}).

\bibitem[{\citenamefont{Liu et~al.}(2013)\citenamefont{Liu, Shan, Yao, Yao, and
  Xiao}}]{ref12}
\bibinfo{author}{\bibfnamefont{G.-B.} \bibnamefont{Liu}},
  \bibinfo{author}{\bibfnamefont{W.-Y.} \bibnamefont{Shan}},
  \bibinfo{author}{\bibfnamefont{Y.}~\bibnamefont{Yao}},
  \bibinfo{author}{\bibfnamefont{W.}~\bibnamefont{Yao}}, \bibnamefont{and}
  \bibinfo{author}{\bibfnamefont{D.}~\bibnamefont{Xiao}},
  \bibinfo{journal}{Phys. Rev. B} \textbf{\bibinfo{volume}{88}},
  \bibinfo{pages}{085433} (\bibinfo{year}{2013}).

\bibitem[{\citenamefont{Leb\`egue and Eriksson}(2009)}]{ref13}
\bibinfo{author}{\bibfnamefont{S.}~\bibnamefont{Leb\`egue}} \bibnamefont{and}
  \bibinfo{author}{\bibfnamefont{O.}~\bibnamefont{Eriksson}},
  \bibinfo{journal}{Phys. Rev. B} \textbf{\bibinfo{volume}{79}},
  \bibinfo{pages}{115409} (\bibinfo{year}{2009}).

\bibitem[{\citenamefont{Zhu et~al.}(2015)\citenamefont{Zhu, Chen, and
  Cui}}]{ref010}
\bibinfo{author}{\bibfnamefont{B.}~\bibnamefont{Zhu}},
  \bibinfo{author}{\bibfnamefont{X.}~\bibnamefont{Chen}}, \bibnamefont{and}
  \bibinfo{author}{\bibfnamefont{X.}~\bibnamefont{Cui}},
  \bibinfo{journal}{Scientific reports} \textbf{\bibinfo{volume}{5}},
  \bibinfo{pages}{9218} (\bibinfo{year}{2015}).

\bibitem[{\citenamefont{Kidd et~al.}(2016)\citenamefont{Kidd, Zhang, and
  Varga}}]{ref011}
\bibinfo{author}{\bibfnamefont{D.~W.} \bibnamefont{Kidd}},
  \bibinfo{author}{\bibfnamefont{D.~K.} \bibnamefont{Zhang}}, \bibnamefont{and}
  \bibinfo{author}{\bibfnamefont{K.}~\bibnamefont{Varga}},
  \bibinfo{journal}{Physical Review B} \textbf{\bibinfo{volume}{93}},
  \bibinfo{pages}{125423} (\bibinfo{year}{2016}).

\bibitem[{\citenamefont{Kyl{\"a}np{\"a}{\"a} and Komsa}(2015)}]{ref012}
\bibinfo{author}{\bibfnamefont{I.}~\bibnamefont{Kyl{\"a}np{\"a}{\"a}}}
  \bibnamefont{and} \bibinfo{author}{\bibfnamefont{H.-P.} \bibnamefont{Komsa}},
  \bibinfo{journal}{Physical Review B} \textbf{\bibinfo{volume}{92}},
  \bibinfo{pages}{205418} (\bibinfo{year}{2015}).

\bibitem[{\citenamefont{Raja et~al.}(2017)\citenamefont{Raja, Chaves, Yu,
  Arefe, Hill, Rigosi, Berkelbach, Nagler, Sch{\"u}ller, Korn et~al.}}]{ref013}
\bibinfo{author}{\bibfnamefont{A.}~\bibnamefont{Raja}},
  \bibinfo{author}{\bibfnamefont{A.}~\bibnamefont{Chaves}},
  \bibinfo{author}{\bibfnamefont{J.}~\bibnamefont{Yu}},
  \bibinfo{author}{\bibfnamefont{G.}~\bibnamefont{Arefe}},
  \bibinfo{author}{\bibfnamefont{H.~M.} \bibnamefont{Hill}},
  \bibinfo{author}{\bibfnamefont{A.~F.} \bibnamefont{Rigosi}},
  \bibinfo{author}{\bibfnamefont{T.~C.} \bibnamefont{Berkelbach}},
  \bibinfo{author}{\bibfnamefont{P.}~\bibnamefont{Nagler}},
  \bibinfo{author}{\bibfnamefont{C.}~\bibnamefont{Sch{\"u}ller}},
  \bibinfo{author}{\bibfnamefont{T.}~\bibnamefont{Korn}}, \bibnamefont{et~al.},
  \bibinfo{journal}{Nature communications} \textbf{\bibinfo{volume}{8}},
  \bibinfo{pages}{15251} (\bibinfo{year}{2017}).

\bibitem[{\citenamefont{Stier et~al.}(2016)\citenamefont{Stier, Wilson, Clark,
  Xu, and Crooker}}]{ref014}
\bibinfo{author}{\bibfnamefont{A.~V.} \bibnamefont{Stier}},
  \bibinfo{author}{\bibfnamefont{N.~P.} \bibnamefont{Wilson}},
  \bibinfo{author}{\bibfnamefont{G.}~\bibnamefont{Clark}},
  \bibinfo{author}{\bibfnamefont{X.}~\bibnamefont{Xu}}, \bibnamefont{and}
  \bibinfo{author}{\bibfnamefont{S.~A.} \bibnamefont{Crooker}},
  \bibinfo{journal}{Nano letters} \textbf{\bibinfo{volume}{16}},
  \bibinfo{pages}{7054} (\bibinfo{year}{2016}).

\bibitem[{\citenamefont{Wang et~al.}(2012)\citenamefont{Wang, Kalantar-Zadeh,
  Kis, Coleman, and Strano}}]{ref05}
\bibinfo{author}{\bibfnamefont{Q.~H.} \bibnamefont{Wang}},
  \bibinfo{author}{\bibfnamefont{K.}~\bibnamefont{Kalantar-Zadeh}},
  \bibinfo{author}{\bibfnamefont{A.}~\bibnamefont{Kis}},
  \bibinfo{author}{\bibfnamefont{J.~N.} \bibnamefont{Coleman}},
  \bibnamefont{and} \bibinfo{author}{\bibfnamefont{M.~S.}
  \bibnamefont{Strano}}, \bibinfo{journal}{Nature nanotechnology}
  \textbf{\bibinfo{volume}{7}}, \bibinfo{pages}{699} (\bibinfo{year}{2012}).

\bibitem[{\citenamefont{Choi et~al.}(2014)\citenamefont{Choi, Qu, Lee, Liu,
  Watanabe, Taniguchi, and Yoo}}]{ref07}
\bibinfo{author}{\bibfnamefont{M.~S.} \bibnamefont{Choi}},
  \bibinfo{author}{\bibfnamefont{D.}~\bibnamefont{Qu}},
  \bibinfo{author}{\bibfnamefont{D.}~\bibnamefont{Lee}},
  \bibinfo{author}{\bibfnamefont{X.}~\bibnamefont{Liu}},
  \bibinfo{author}{\bibfnamefont{K.}~\bibnamefont{Watanabe}},
  \bibinfo{author}{\bibfnamefont{T.}~\bibnamefont{Taniguchi}},
  \bibnamefont{and} \bibinfo{author}{\bibfnamefont{W.~J.} \bibnamefont{Yoo}},
  \bibinfo{journal}{ACS nano} \textbf{\bibinfo{volume}{8}},
  \bibinfo{pages}{9332} (\bibinfo{year}{2014}).

\bibitem[{\citenamefont{Zhang et~al.}(2015)\citenamefont{Zhang, Feng, Wang,
  Azcatl, Lu, Addou, Wang, Zhou, Lerach, Bojan et~al.}}]{ref017}
\bibinfo{author}{\bibfnamefont{K.}~\bibnamefont{Zhang}},
  \bibinfo{author}{\bibfnamefont{S.}~\bibnamefont{Feng}},
  \bibinfo{author}{\bibfnamefont{J.}~\bibnamefont{Wang}},
  \bibinfo{author}{\bibfnamefont{A.}~\bibnamefont{Azcatl}},
  \bibinfo{author}{\bibfnamefont{N.}~\bibnamefont{Lu}},
  \bibinfo{author}{\bibfnamefont{R.}~\bibnamefont{Addou}},
  \bibinfo{author}{\bibfnamefont{N.}~\bibnamefont{Wang}},
  \bibinfo{author}{\bibfnamefont{C.}~\bibnamefont{Zhou}},
  \bibinfo{author}{\bibfnamefont{J.}~\bibnamefont{Lerach}},
  \bibinfo{author}{\bibfnamefont{V.}~\bibnamefont{Bojan}},
  \bibnamefont{et~al.}, \bibinfo{journal}{Nano letters}
  \textbf{\bibinfo{volume}{15}}, \bibinfo{pages}{6586} (\bibinfo{year}{2015}).

\bibitem[{\citenamefont{Sundaram et~al.}(2013)\citenamefont{Sundaram, Engel,
  Lombardo, Krupke, Ferrari, Avouris, and Steiner}}]{ref018}
\bibinfo{author}{\bibfnamefont{R.}~\bibnamefont{Sundaram}},
  \bibinfo{author}{\bibfnamefont{M.}~\bibnamefont{Engel}},
  \bibinfo{author}{\bibfnamefont{A.}~\bibnamefont{Lombardo}},
  \bibinfo{author}{\bibfnamefont{R.}~\bibnamefont{Krupke}},
  \bibinfo{author}{\bibfnamefont{A.}~\bibnamefont{Ferrari}},
  \bibinfo{author}{\bibfnamefont{P.}~\bibnamefont{Avouris}}, \bibnamefont{and}
  \bibinfo{author}{\bibfnamefont{M.}~\bibnamefont{Steiner}},
  \bibinfo{journal}{Nano letters} \textbf{\bibinfo{volume}{13}},
  \bibinfo{pages}{1416} (\bibinfo{year}{2013}).

\bibitem[{\citenamefont{Ye et~al.}(2014)\citenamefont{Ye, Ye, Gharghi, Zhu,
  Zhao, Wang, Yin, and Zhang}}]{ref019}
\bibinfo{author}{\bibfnamefont{Y.}~\bibnamefont{Ye}},
  \bibinfo{author}{\bibfnamefont{Z.}~\bibnamefont{Ye}},
  \bibinfo{author}{\bibfnamefont{M.}~\bibnamefont{Gharghi}},
  \bibinfo{author}{\bibfnamefont{H.}~\bibnamefont{Zhu}},
  \bibinfo{author}{\bibfnamefont{M.}~\bibnamefont{Zhao}},
  \bibinfo{author}{\bibfnamefont{Y.}~\bibnamefont{Wang}},
  \bibinfo{author}{\bibfnamefont{X.}~\bibnamefont{Yin}}, \bibnamefont{and}
  \bibinfo{author}{\bibfnamefont{X.}~\bibnamefont{Zhang}},
  \bibinfo{journal}{Applied Physics Letters} \textbf{\bibinfo{volume}{104}},
  \bibinfo{pages}{193508} (\bibinfo{year}{2014}).

\bibitem[{\citenamefont{Ye et~al.}(2015)\citenamefont{Ye, Wong, Lu, Ni, Zhu,
  Chen, Wang, and Zhang}}]{ref020}
\bibinfo{author}{\bibfnamefont{Y.}~\bibnamefont{Ye}},
  \bibinfo{author}{\bibfnamefont{Z.~J.} \bibnamefont{Wong}},
  \bibinfo{author}{\bibfnamefont{X.}~\bibnamefont{Lu}},
  \bibinfo{author}{\bibfnamefont{X.}~\bibnamefont{Ni}},
  \bibinfo{author}{\bibfnamefont{H.}~\bibnamefont{Zhu}},
  \bibinfo{author}{\bibfnamefont{X.}~\bibnamefont{Chen}},
  \bibinfo{author}{\bibfnamefont{Y.}~\bibnamefont{Wang}}, \bibnamefont{and}
  \bibinfo{author}{\bibfnamefont{X.}~\bibnamefont{Zhang}},
  \bibinfo{journal}{Nature Photonics} \textbf{\bibinfo{volume}{9}},
  \bibinfo{pages}{733} (\bibinfo{year}{2015}).

\bibitem[{\citenamefont{Cheng et~al.}(2013)\citenamefont{Cheng, Zhu, Mi, Guo,
  and Schwingenschl\"ogl}}]{ref6}
\bibinfo{author}{\bibfnamefont{Y.~C.} \bibnamefont{Cheng}},
  \bibinfo{author}{\bibfnamefont{Z.~Y.} \bibnamefont{Zhu}},
  \bibinfo{author}{\bibfnamefont{W.~B.} \bibnamefont{Mi}},
  \bibinfo{author}{\bibfnamefont{Z.~B.} \bibnamefont{Guo}}, \bibnamefont{and}
  \bibinfo{author}{\bibfnamefont{U.}~\bibnamefont{Schwingenschl\"ogl}},
  \bibinfo{journal}{Phys. Rev. B} \textbf{\bibinfo{volume}{87}},
  \bibinfo{pages}{100401} (\bibinfo{year}{2013}).

\bibitem[{\citenamefont{Mishra et~al.}(2013)\citenamefont{Mishra, Zhou,
  Pennycook, Pantelides, and Idrobo}}]{ref7}
\bibinfo{author}{\bibfnamefont{R.}~\bibnamefont{Mishra}},
  \bibinfo{author}{\bibfnamefont{W.}~\bibnamefont{Zhou}},
  \bibinfo{author}{\bibfnamefont{S.~J.} \bibnamefont{Pennycook}},
  \bibinfo{author}{\bibfnamefont{S.~T.} \bibnamefont{Pantelides}},
  \bibnamefont{and} \bibinfo{author}{\bibfnamefont{J.-C.}
  \bibnamefont{Idrobo}}, \bibinfo{journal}{Phys. Rev. B}
  \textbf{\bibinfo{volume}{88}}, \bibinfo{pages}{144409}
  (\bibinfo{year}{2013}).

\bibitem[{\citenamefont{Fang et~al.}(2018)\citenamefont{Fang, Zhao, Huang, Xu,
  Min, Chu, and Ma}}]{ref8}
\bibinfo{author}{\bibfnamefont{Q.}~\bibnamefont{Fang}},
  \bibinfo{author}{\bibfnamefont{X.}~\bibnamefont{Zhao}},
  \bibinfo{author}{\bibfnamefont{Y.}~\bibnamefont{Huang}},
  \bibinfo{author}{\bibfnamefont{K.}~\bibnamefont{Xu}},
  \bibinfo{author}{\bibfnamefont{T.}~\bibnamefont{Min}},
  \bibinfo{author}{\bibfnamefont{P.~K.} \bibnamefont{Chu}}, \bibnamefont{and}
  \bibinfo{author}{\bibfnamefont{F.}~\bibnamefont{Ma}}, \bibinfo{journal}{Phys.
  Chem. Chem. Phys.} \textbf{\bibinfo{volume}{20}}, \bibinfo{pages}{553}
  (\bibinfo{year}{2018}).

\bibitem[{\citenamefont{Ramasubramaniam and Naveh}(2013)}]{ref9}
\bibinfo{author}{\bibfnamefont{A.}~\bibnamefont{Ramasubramaniam}}
  \bibnamefont{and} \bibinfo{author}{\bibfnamefont{D.}~\bibnamefont{Naveh}},
  \bibinfo{journal}{Phys. Rev. B} \textbf{\bibinfo{volume}{87}},
  \bibinfo{pages}{195201} (\bibinfo{year}{2013}).

\bibitem[{\citenamefont{Fan et~al.}(2016)\citenamefont{Fan, An, and
  Guo}}]{ref10}
\bibinfo{author}{\bibfnamefont{X.-L.} \bibnamefont{Fan}},
  \bibinfo{author}{\bibfnamefont{Y.-R.} \bibnamefont{An}}, \bibnamefont{and}
  \bibinfo{author}{\bibfnamefont{W.-J.} \bibnamefont{Guo}},
  \bibinfo{journal}{Nanoscale Research Letters} \textbf{\bibinfo{volume}{11}},
  \bibinfo{pages}{154} (\bibinfo{year}{2016}), ISSN \bibinfo{issn}{1556-276X}.

\bibitem[{\citenamefont{Wang et~al.}(2016)\citenamefont{Wang, Sun, Yang, Li,
  Zhao, Xu, Zhang, and Zeng}}]{ref14}
\bibinfo{author}{\bibfnamefont{J.}~\bibnamefont{Wang}},
  \bibinfo{author}{\bibfnamefont{F.}~\bibnamefont{Sun}},
  \bibinfo{author}{\bibfnamefont{S.}~\bibnamefont{Yang}},
  \bibinfo{author}{\bibfnamefont{Y.}~\bibnamefont{Li}},
  \bibinfo{author}{\bibfnamefont{C.}~\bibnamefont{Zhao}},
  \bibinfo{author}{\bibfnamefont{M.}~\bibnamefont{Xu}},
  \bibinfo{author}{\bibfnamefont{Y.}~\bibnamefont{Zhang}}, \bibnamefont{and}
  \bibinfo{author}{\bibfnamefont{H.}~\bibnamefont{Zeng}},
  \bibinfo{journal}{Applied Physics Letters} \textbf{\bibinfo{volume}{109}},
  \bibinfo{pages}{092401} (\bibinfo{year}{2016}).

\bibitem[{\citenamefont{Cong et~al.}(2015)\citenamefont{Cong, Tang, Zhao, and
  Chu}}]{ref047}
\bibinfo{author}{\bibfnamefont{W.}~\bibnamefont{Cong}},
  \bibinfo{author}{\bibfnamefont{Z.}~\bibnamefont{Tang}},
  \bibinfo{author}{\bibfnamefont{X.}~\bibnamefont{Zhao}}, \bibnamefont{and}
  \bibinfo{author}{\bibfnamefont{J.}~\bibnamefont{Chu}},
  \bibinfo{journal}{Scientific reports} \textbf{\bibinfo{volume}{5}},
  \bibinfo{pages}{9361} (\bibinfo{year}{2015}).

\bibitem[{\citenamefont{Yue et~al.}(2013)\citenamefont{Yue, Chang, Qin, and
  Li}}]{ref060}
\bibinfo{author}{\bibfnamefont{Q.}~\bibnamefont{Yue}},
  \bibinfo{author}{\bibfnamefont{S.}~\bibnamefont{Chang}},
  \bibinfo{author}{\bibfnamefont{S.}~\bibnamefont{Qin}}, \bibnamefont{and}
  \bibinfo{author}{\bibfnamefont{J.}~\bibnamefont{Li}},
  \bibinfo{journal}{Physics Letters A} \textbf{\bibinfo{volume}{377}},
  \bibinfo{pages}{1362} (\bibinfo{year}{2013}).

\bibitem[{\citenamefont{Song et~al.}(2017)\citenamefont{Song, Tong, Shen, Gong,
  Tang, and Duan}}]{ref048}
\bibinfo{author}{\bibfnamefont{Y.-X.} \bibnamefont{Song}},
  \bibinfo{author}{\bibfnamefont{W.-Y.} \bibnamefont{Tong}},
  \bibinfo{author}{\bibfnamefont{Y.-H.} \bibnamefont{Shen}},
  \bibinfo{author}{\bibfnamefont{S.-J.} \bibnamefont{Gong}},
  \bibinfo{author}{\bibfnamefont{Z.}~\bibnamefont{Tang}}, \bibnamefont{and}
  \bibinfo{author}{\bibfnamefont{C.-G.} \bibnamefont{Duan}},
  \bibinfo{journal}{Journal of Physics: Condensed Matter}
  \textbf{\bibinfo{volume}{29}}, \bibinfo{pages}{475803}
  (\bibinfo{year}{2017}).

\bibitem[{\citenamefont{Sun et~al.}(2016)\citenamefont{Sun, Zhou, Liang, Liu,
  and Wu}}]{ref015}
\bibinfo{author}{\bibfnamefont{L.}~\bibnamefont{Sun}},
  \bibinfo{author}{\bibfnamefont{W.}~\bibnamefont{Zhou}},
  \bibinfo{author}{\bibfnamefont{Y.}~\bibnamefont{Liang}},
  \bibinfo{author}{\bibfnamefont{L.}~\bibnamefont{Liu}}, \bibnamefont{and}
  \bibinfo{author}{\bibfnamefont{P.}~\bibnamefont{Wu}},
  \bibinfo{journal}{Computational Materials Science}
  \textbf{\bibinfo{volume}{117}}, \bibinfo{pages}{489} (\bibinfo{year}{2016}).

\bibitem[{\citenamefont{Gao et~al.}(2016)\citenamefont{Gao, Kim, Liang, Idrobo,
  Chow, Tan, Li, Li, Sumpter, Lu et~al.}}]{ref15}
\bibinfo{author}{\bibfnamefont{J.}~\bibnamefont{Gao}},
  \bibinfo{author}{\bibfnamefont{Y.~D.} \bibnamefont{Kim}},
  \bibinfo{author}{\bibfnamefont{L.}~\bibnamefont{Liang}},
  \bibinfo{author}{\bibfnamefont{J.~C.} \bibnamefont{Idrobo}},
  \bibinfo{author}{\bibfnamefont{P.}~\bibnamefont{Chow}},
  \bibinfo{author}{\bibfnamefont{J.}~\bibnamefont{Tan}},
  \bibinfo{author}{\bibfnamefont{B.}~\bibnamefont{Li}},
  \bibinfo{author}{\bibfnamefont{L.}~\bibnamefont{Li}},
  \bibinfo{author}{\bibfnamefont{B.~G.} \bibnamefont{Sumpter}},
  \bibinfo{author}{\bibfnamefont{T.-M.} \bibnamefont{Lu}},
  \bibnamefont{et~al.}, \bibinfo{journal}{Advanced Materials}
  \textbf{\bibinfo{volume}{28}}, \bibinfo{pages}{9735} (\bibinfo{year}{2016}).

\bibitem[{\citenamefont{Seixas et~al.}(2015)\citenamefont{Seixas, Carvalho, and
  Neto}}]{ref016}
\bibinfo{author}{\bibfnamefont{L.}~\bibnamefont{Seixas}},
  \bibinfo{author}{\bibfnamefont{A.}~\bibnamefont{Carvalho}}, \bibnamefont{and}
  \bibinfo{author}{\bibfnamefont{A.~C.} \bibnamefont{Neto}},
  \bibinfo{journal}{Physical Review B} \textbf{\bibinfo{volume}{91}},
  \bibinfo{pages}{155138} (\bibinfo{year}{2015}).

\bibitem[{\citenamefont{Cai et~al.}(2014)\citenamefont{Cai, Zhang, and
  Zhang}}]{ref049}
\bibinfo{author}{\bibfnamefont{Y.}~\bibnamefont{Cai}},
  \bibinfo{author}{\bibfnamefont{G.}~\bibnamefont{Zhang}}, \bibnamefont{and}
  \bibinfo{author}{\bibfnamefont{Y.-W.} \bibnamefont{Zhang}},
  \bibinfo{journal}{Journal of the American Chemical Society}
  \textbf{\bibinfo{volume}{136}}, \bibinfo{pages}{6269} (\bibinfo{year}{2014}).

\bibitem[{\citenamefont{Baugher et~al.}(2013)\citenamefont{Baugher, Churchill,
  Yang, and Jarillo-Herrero}}]{ref050}
\bibinfo{author}{\bibfnamefont{B.~W.} \bibnamefont{Baugher}},
  \bibinfo{author}{\bibfnamefont{H.~O.} \bibnamefont{Churchill}},
  \bibinfo{author}{\bibfnamefont{Y.}~\bibnamefont{Yang}}, \bibnamefont{and}
  \bibinfo{author}{\bibfnamefont{P.}~\bibnamefont{Jarillo-Herrero}},
  \bibinfo{journal}{Nano letters} \textbf{\bibinfo{volume}{13}},
  \bibinfo{pages}{4212} (\bibinfo{year}{2013}).

\bibitem[{\citenamefont{Lembke and Kis}(2012)}]{ref051}
\bibinfo{author}{\bibfnamefont{D.}~\bibnamefont{Lembke}} \bibnamefont{and}
  \bibinfo{author}{\bibfnamefont{A.}~\bibnamefont{Kis}}, \bibinfo{journal}{ACS
  nano} \textbf{\bibinfo{volume}{6}}, \bibinfo{pages}{10070}
  (\bibinfo{year}{2012}).

\bibitem[{\citenamefont{Schmidt et~al.}(2014)\citenamefont{Schmidt, Wang, Chu,
  Toh, Kumar, Zhao, Castro~Neto, Martin, Adam, Özyilmaz et~al.}}]{ref054}
\bibinfo{author}{\bibfnamefont{H.}~\bibnamefont{Schmidt}},
  \bibinfo{author}{\bibfnamefont{S.}~\bibnamefont{Wang}},
  \bibinfo{author}{\bibfnamefont{L.}~\bibnamefont{Chu}},
  \bibinfo{author}{\bibfnamefont{M.}~\bibnamefont{Toh}},
  \bibinfo{author}{\bibfnamefont{R.}~\bibnamefont{Kumar}},
  \bibinfo{author}{\bibfnamefont{W.}~\bibnamefont{Zhao}},
  \bibinfo{author}{\bibfnamefont{A.}~\bibnamefont{Castro~Neto}},
  \bibinfo{author}{\bibfnamefont{J.}~\bibnamefont{Martin}},
  \bibinfo{author}{\bibfnamefont{S.}~\bibnamefont{Adam}},
  \bibinfo{author}{\bibfnamefont{B.}~\bibnamefont{Özyilmaz}},
  \bibnamefont{et~al.}, \bibinfo{journal}{Nano letters}
  \textbf{\bibinfo{volume}{14}}, \bibinfo{pages}{1909} (\bibinfo{year}{2014}).

\bibitem[{\citenamefont{Radisavljevic et~al.}(2011)\citenamefont{Radisavljevic,
  Whitwick, and Kis}}]{ref052}
\bibinfo{author}{\bibfnamefont{B.}~\bibnamefont{Radisavljevic}},
  \bibinfo{author}{\bibfnamefont{M.~B.} \bibnamefont{Whitwick}},
  \bibnamefont{and} \bibinfo{author}{\bibfnamefont{A.}~\bibnamefont{Kis}},
  \bibinfo{journal}{ACS nano} \textbf{\bibinfo{volume}{5}},
  \bibinfo{pages}{9934} (\bibinfo{year}{2011}).

\bibitem[{\citenamefont{Zhang et~al.}(2012)\citenamefont{Zhang, Ye, Matsuhashi,
  and Iwasa}}]{ref053}
\bibinfo{author}{\bibfnamefont{Y.}~\bibnamefont{Zhang}},
  \bibinfo{author}{\bibfnamefont{J.}~\bibnamefont{Ye}},
  \bibinfo{author}{\bibfnamefont{Y.}~\bibnamefont{Matsuhashi}},
  \bibnamefont{and} \bibinfo{author}{\bibfnamefont{Y.}~\bibnamefont{Iwasa}},
  \bibinfo{journal}{Nano letters} \textbf{\bibinfo{volume}{12}},
  \bibinfo{pages}{1136} (\bibinfo{year}{2012}).

\bibitem[{\citenamefont{Lin et~al.}(2013)\citenamefont{Lin, Zhong, Zhong, Li,
  Zhang, and Chen}}]{ref055}
\bibinfo{author}{\bibfnamefont{J.}~\bibnamefont{Lin}},
  \bibinfo{author}{\bibfnamefont{J.}~\bibnamefont{Zhong}},
  \bibinfo{author}{\bibfnamefont{S.}~\bibnamefont{Zhong}},
  \bibinfo{author}{\bibfnamefont{H.}~\bibnamefont{Li}},
  \bibinfo{author}{\bibfnamefont{H.}~\bibnamefont{Zhang}}, \bibnamefont{and}
  \bibinfo{author}{\bibfnamefont{W.}~\bibnamefont{Chen}},
  \bibinfo{journal}{Applied Physics Letters} \textbf{\bibinfo{volume}{103}},
  \bibinfo{pages}{063109} (\bibinfo{year}{2013}).

\bibitem[{\citenamefont{Sato et~al.}(2010)\citenamefont{Sato, Bergqvist,
  Kudrnovsk{\`y}, Dederichs, Eriksson, Turek, Sanyal, Bouzerar,
  Katayama-Yoshida, Dinh et~al.}}]{ref033}
\bibinfo{author}{\bibfnamefont{K.}~\bibnamefont{Sato}},
  \bibinfo{author}{\bibfnamefont{L.}~\bibnamefont{Bergqvist}},
  \bibinfo{author}{\bibfnamefont{J.}~\bibnamefont{Kudrnovsk{\`y}}},
  \bibinfo{author}{\bibfnamefont{P.~H.} \bibnamefont{Dederichs}},
  \bibinfo{author}{\bibfnamefont{O.}~\bibnamefont{Eriksson}},
  \bibinfo{author}{\bibfnamefont{I.}~\bibnamefont{Turek}},
  \bibinfo{author}{\bibfnamefont{B.}~\bibnamefont{Sanyal}},
  \bibinfo{author}{\bibfnamefont{G.}~\bibnamefont{Bouzerar}},
  \bibinfo{author}{\bibfnamefont{H.}~\bibnamefont{Katayama-Yoshida}},
  \bibinfo{author}{\bibfnamefont{V.}~\bibnamefont{Dinh}}, \bibnamefont{et~al.},
  \bibinfo{journal}{Reviews of modern physics} \textbf{\bibinfo{volume}{82}},
  \bibinfo{pages}{1633} (\bibinfo{year}{2010}).

\bibitem[{\citenamefont{Andriotis and Menon}(2014)}]{ref032}
\bibinfo{author}{\bibfnamefont{A.~N.} \bibnamefont{Andriotis}}
  \bibnamefont{and} \bibinfo{author}{\bibfnamefont{M.}~\bibnamefont{Menon}},
  \bibinfo{journal}{Physical Review B} \textbf{\bibinfo{volume}{90}},
  \bibinfo{pages}{125304} (\bibinfo{year}{2014}).

\bibitem[{\citenamefont{Andriotis et~al.}(2015)\citenamefont{Andriotis,
  Fthenakis, and Menon}}]{ref034}
\bibinfo{author}{\bibfnamefont{A.~N.} \bibnamefont{Andriotis}},
  \bibinfo{author}{\bibfnamefont{Z.~G.} \bibnamefont{Fthenakis}},
  \bibnamefont{and} \bibinfo{author}{\bibfnamefont{M.}~\bibnamefont{Menon}},
  \bibinfo{journal}{Journal of Physics: Condensed Matter}
  \textbf{\bibinfo{volume}{27}}, \bibinfo{pages}{052202}
  (\bibinfo{year}{2015}).

\bibitem[{\citenamefont{Andriotis and Menon}(2012)}]{ref035}
\bibinfo{author}{\bibfnamefont{A.~N.} \bibnamefont{Andriotis}}
  \bibnamefont{and} \bibinfo{author}{\bibfnamefont{M.}~\bibnamefont{Menon}},
  \bibinfo{journal}{Journal of Physics: Condensed Matter}
  \textbf{\bibinfo{volume}{24}}, \bibinfo{pages}{455801}
  (\bibinfo{year}{2012}).

\bibitem[{\citenamefont{Andriotis and Menon}(2013)}]{ref036}
\bibinfo{author}{\bibfnamefont{A.~N.} \bibnamefont{Andriotis}}
  \bibnamefont{and} \bibinfo{author}{\bibfnamefont{M.}~\bibnamefont{Menon}},
  \bibinfo{journal}{Physical Review B} \textbf{\bibinfo{volume}{87}},
  \bibinfo{pages}{155309} (\bibinfo{year}{2013}).

\bibitem[{\citenamefont{Lin and Ni}(2016)}]{ref056}
\bibinfo{author}{\bibfnamefont{X.}~\bibnamefont{Lin}} \bibnamefont{and}
  \bibinfo{author}{\bibfnamefont{J.}~\bibnamefont{Ni}},
  \bibinfo{journal}{Journal of Applied Physics} \textbf{\bibinfo{volume}{120}},
  \bibinfo{pages}{064305} (\bibinfo{year}{2016}).

\bibitem[{\citenamefont{Lisenkov et~al.}(2012)\citenamefont{Lisenkov,
  Andriotis, and Menon}}]{ref057}
\bibinfo{author}{\bibfnamefont{S.}~\bibnamefont{Lisenkov}},
  \bibinfo{author}{\bibfnamefont{A.~N.} \bibnamefont{Andriotis}},
  \bibnamefont{and} \bibinfo{author}{\bibfnamefont{M.}~\bibnamefont{Menon}},
  \bibinfo{journal}{Physical review letters} \textbf{\bibinfo{volume}{108}},
  \bibinfo{pages}{187208} (\bibinfo{year}{2012}).

\bibitem[{\citenamefont{Kresse and Hafner}(1993)}]{ref1}
\bibinfo{author}{\bibfnamefont{G.}~\bibnamefont{Kresse}} \bibnamefont{and}
  \bibinfo{author}{\bibfnamefont{J.}~\bibnamefont{Hafner}},
  \bibinfo{journal}{Phys. Rev. B} \textbf{\bibinfo{volume}{47}},
  \bibinfo{pages}{558} (\bibinfo{year}{1993}).

\bibitem[{\citenamefont{Kresse and Hafner}(1994)}]{ref2}
\bibinfo{author}{\bibfnamefont{G.}~\bibnamefont{Kresse}} \bibnamefont{and}
  \bibinfo{author}{\bibfnamefont{J.}~\bibnamefont{Hafner}},
  \bibinfo{journal}{Phys. Rev. B} \textbf{\bibinfo{volume}{49}},
  \bibinfo{pages}{14251} (\bibinfo{year}{1994}).

\bibitem[{\citenamefont{Perdew et~al.}(1996)\citenamefont{Perdew, Burke, and
  Ernzerhof}}]{ref4}
\bibinfo{author}{\bibfnamefont{J.~P.} \bibnamefont{Perdew}},
  \bibinfo{author}{\bibfnamefont{K.}~\bibnamefont{Burke}}, \bibnamefont{and}
  \bibinfo{author}{\bibfnamefont{M.}~\bibnamefont{Ernzerhof}},
  \bibinfo{journal}{Phys. Rev. Lett.} \textbf{\bibinfo{volume}{77}},
  \bibinfo{pages}{3865} (\bibinfo{year}{1996}).

\bibitem[{\citenamefont{Dudarev et~al.}(1998)\citenamefont{Dudarev, Botton,
  Savrasov, Humphreys, and Sutton}}]{DFTU}
\bibinfo{author}{\bibfnamefont{S.~L.} \bibnamefont{Dudarev}},
  \bibinfo{author}{\bibfnamefont{G.~A.} \bibnamefont{Botton}},
  \bibinfo{author}{\bibfnamefont{S.~Y.} \bibnamefont{Savrasov}},
  \bibinfo{author}{\bibfnamefont{C.~J.} \bibnamefont{Humphreys}},
  \bibnamefont{and} \bibinfo{author}{\bibfnamefont{A.~P.}
  \bibnamefont{Sutton}}, \bibinfo{journal}{Phys. Rev. B}
  \textbf{\bibinfo{volume}{57}}, \bibinfo{pages}{1505} (\bibinfo{year}{1998}).

\bibitem[{\citenamefont{Wang et~al.}(2017)\citenamefont{Wang, Tseng, Murmu,
  Bao, Kennedy, Ionesc, Ding, Suzuki, Li, and Yi}}]{ref031}
\bibinfo{author}{\bibfnamefont{Y.}~\bibnamefont{Wang}},
  \bibinfo{author}{\bibfnamefont{L.-T.} \bibnamefont{Tseng}},
  \bibinfo{author}{\bibfnamefont{P.~P.} \bibnamefont{Murmu}},
  \bibinfo{author}{\bibfnamefont{N.}~\bibnamefont{Bao}},
  \bibinfo{author}{\bibfnamefont{J.}~\bibnamefont{Kennedy}},
  \bibinfo{author}{\bibfnamefont{M.}~\bibnamefont{Ionesc}},
  \bibinfo{author}{\bibfnamefont{J.}~\bibnamefont{Ding}},
  \bibinfo{author}{\bibfnamefont{K.}~\bibnamefont{Suzuki}},
  \bibinfo{author}{\bibfnamefont{S.}~\bibnamefont{Li}}, \bibnamefont{and}
  \bibinfo{author}{\bibfnamefont{J.}~\bibnamefont{Yi}},
  \bibinfo{journal}{Materials \& Design} \textbf{\bibinfo{volume}{121}},
  \bibinfo{pages}{77} (\bibinfo{year}{2017}).

\bibitem[{\citenamefont{Bl\"ochl}(1994)}]{ref3}
\bibinfo{author}{\bibfnamefont{P.~E.} \bibnamefont{Bl\"ochl}},
  \bibinfo{journal}{Phys. Rev. B} \textbf{\bibinfo{volume}{50}},
  \bibinfo{pages}{17953} (\bibinfo{year}{1994}).

\bibitem[{\citenamefont{Kudrnovsk\'y et~al.}(2004)\citenamefont{Kudrnovsk\'y,
  Turek, Drchal, M\'aca, Weinberger, and Bruno}}]{ref037}
\bibinfo{author}{\bibfnamefont{J.}~\bibnamefont{Kudrnovsk\'y}},
  \bibinfo{author}{\bibfnamefont{I.}~\bibnamefont{Turek}},
  \bibinfo{author}{\bibfnamefont{V.}~\bibnamefont{Drchal}},
  \bibinfo{author}{\bibfnamefont{F.}~\bibnamefont{M\'aca}},
  \bibinfo{author}{\bibfnamefont{P.}~\bibnamefont{Weinberger}},
  \bibnamefont{and} \bibinfo{author}{\bibfnamefont{P.}~\bibnamefont{Bruno}},
  \bibinfo{journal}{Phys. Rev. B} \textbf{\bibinfo{volume}{69}},
  \bibinfo{pages}{115208} (\bibinfo{year}{2004}).

\bibitem[{\citenamefont{Kan et~al.}(2013)\citenamefont{Kan, Zhou, Sun, Kawazoe,
  and Jena}}]{ref038}
\bibinfo{author}{\bibfnamefont{M.}~\bibnamefont{Kan}},
  \bibinfo{author}{\bibfnamefont{J.}~\bibnamefont{Zhou}},
  \bibinfo{author}{\bibfnamefont{Q.}~\bibnamefont{Sun}},
  \bibinfo{author}{\bibfnamefont{Y.}~\bibnamefont{Kawazoe}}, \bibnamefont{and}
  \bibinfo{author}{\bibfnamefont{P.}~\bibnamefont{Jena}}, \bibinfo{journal}{The
  journal of physical chemistry letters} \textbf{\bibinfo{volume}{4}},
  \bibinfo{pages}{3382} (\bibinfo{year}{2013}).

\bibitem[{\citenamefont{Zhou et~al.}(2010)\citenamefont{Zhou, Wiebe, Lounis,
  Vedmedenko, Meier, Bl{\"u}gel, Dederichs, and Wiesendanger}}]{lounis1}
\bibinfo{author}{\bibfnamefont{L.}~\bibnamefont{Zhou}},
  \bibinfo{author}{\bibfnamefont{J.}~\bibnamefont{Wiebe}},
  \bibinfo{author}{\bibfnamefont{S.}~\bibnamefont{Lounis}},
  \bibinfo{author}{\bibfnamefont{E.}~\bibnamefont{Vedmedenko}},
  \bibinfo{author}{\bibfnamefont{F.}~\bibnamefont{Meier}},
  \bibinfo{author}{\bibfnamefont{S.}~\bibnamefont{Bl{\"u}gel}},
  \bibinfo{author}{\bibfnamefont{P.~H.} \bibnamefont{Dederichs}},
  \bibnamefont{and}
  \bibinfo{author}{\bibfnamefont{R.}~\bibnamefont{Wiesendanger}},
  \bibinfo{journal}{Nature Physics} \textbf{\bibinfo{volume}{6}},
  \bibinfo{pages}{187} (\bibinfo{year}{2010}).

\bibitem[{\citenamefont{Loth et~al.}(2012)\citenamefont{Loth, Baumann, Lutz,
  Eigler, and Heinrich}}]{lounis2}
\bibinfo{author}{\bibfnamefont{S.}~\bibnamefont{Loth}},
  \bibinfo{author}{\bibfnamefont{S.}~\bibnamefont{Baumann}},
  \bibinfo{author}{\bibfnamefont{C.~P.} \bibnamefont{Lutz}},
  \bibinfo{author}{\bibfnamefont{D.}~\bibnamefont{Eigler}}, \bibnamefont{and}
  \bibinfo{author}{\bibfnamefont{A.~J.} \bibnamefont{Heinrich}},
  \bibinfo{journal}{Science} \textbf{\bibinfo{volume}{335}},
  \bibinfo{pages}{196} (\bibinfo{year}{2012}).

\end{thebibliography}
%%%%%%%%%%%%%%%%%%%%%%%%%%%%%%%%%%%%%%%%%%%%%%%%%%%%%%%%%%%%%%%%%%%%%%%%%%%%%%%%%%%
\end{document}